\documentclass[aps,prl,twocolumn,amsmath,amssymb,superscriptaddress,nofootinbib,showpacs,preprintnumbers]{revtex4-1}

\usepackage{graphicx}

\graphicspath{{Figures/}}
\def\lf{\left\lfloor}   
\def\rf{\right\rfloor}

\usepackage{bm}
\usepackage{amsmath,amssymb}
\usepackage{aas_macros}
\usepackage{natbib}
\usepackage[colorlinks=true]{hyperref}

\begin{document}


\title{Searching for dark matter with an unequal delay interferometer}

\author{Etienne Savalle}
\affiliation{SYRTE, Observatoire de Paris, Universit\'e PSL, CNRS, Sorbonne Universit\'e, LNE, 75014 Paris, France}
\author{Aur\'elien Hees}
\affiliation{SYRTE, Observatoire de Paris, Universit\'e PSL, CNRS, Sorbonne Universit\'e, LNE, 75014 Paris, France}
\author{Florian Frank}
\affiliation{SYRTE, Observatoire de Paris, Universit\'e PSL, CNRS, Sorbonne Universit\'e, LNE, 75014 Paris, France}
\author{Etienne Cantin}
\affiliation{SYRTE, Observatoire de Paris, Universit\'e PSL, CNRS, Sorbonne Universit\'e, LNE, 75014 Paris, France}
\author{Paul-Eric Pottie}
\affiliation{SYRTE, Observatoire de Paris, Universit\'e PSL, CNRS, Sorbonne Universit\'e, LNE, 75014 Paris, France}
\author{Benjamin M. Roberts}
\affiliation{School of Mathematics and Physics, The University of Queensland, Brisbane QLD 4072, Australia}
\author{Lucie Cros}
\affiliation{SYRTE, Observatoire de Paris, Universit\'e PSL, CNRS, Sorbonne Universit\'e, LNE, 75014 Paris, France}
\affiliation{MINES ParisTech, Universit\'e PSL, 75006 Paris, France}
\author{Ben T. McAllister}
\affiliation{ARC Centre of Excellence for Engineered Quantum Systems, School of Physics, University of Western Australia, Crawley WA 6009, Australia}
\author{Peter Wolf}
\affiliation{SYRTE, Observatoire de Paris, Universit\'e PSL, CNRS, Sorbonne Universit\'e, LNE, 75014 Paris, France}
\email{peter.wolf@obspm.fr}

\date{\today}
\begin{abstract}
We propose a new type of experiment that compares the frequency of a clock (an ultra-stable optical cavity in this case) at time $t$ to its own frequency some time $t-T$ earlier, by ``storing" the output signal (photons) in a fibre delay line. In ultra-light oscillating dark matter (DM) models, such an experiment is sensitive to coupling of DM to the standard model fields, through oscillations of the cavity and fibre lengths and of the fibre refractive index. Additionally, the sensitivity is significantly enhanced around the mechanical resonances of the cavity. We present experimental result of such an experiment and report no evidence of DM for masses in the [$4.1\times 10^{-11}$, $8.3\times 10^{-10}$]~eV region. In addition, we improve constraints on the involved coupling constants by one order of magnitude in a standard galactic DM model, at the mass corresponding to the resonant frequency of our cavity. Furthermore, in the model of relaxion DM, we improve on existing constraints over the whole DM mass range by about one order of magnitude, and up to six orders of magnitude at resonance.     
\end{abstract}

\maketitle

\paragraph{\textbf{Introduction}}

Dark matter (DM) remains one of the contemporary mysteries in fundamental physics and continues to question the scientific community regarding its origin and composition. DM leaves indirect evidences of its existence through its gravitational interaction but has never been directly detected so far \cite{Bertone2018}, leading to the development of a multitude of experiments probing different models covering a large mass range \cite{Bertone2018,AtomicReview2017}.  

Amongst the various DM models, ultra-light DM scenarios have recently seen a strong surge thanks to the excellent sensitivities provided by the latest advances in time/frequency metrology 
\cite{DereviankoDM2014,Arvanitaki2014,StadnikDMalpha2015,Tilburg2015,
Hees2016,Wcislo2016,GPSDM2017,Hees2018,Wcislo2018,RobertsAsymm2018,
GPSDM2018,Alonso2019,Wolf2019,Kennedy2020}. In such class of models, DM is made of a scalar-field (SF) non-universally coupled to the Standard Model. Typically, this SF will undergo oscillations that will be reflected through an oscillation of the fundamental constants of Nature, a signature of a violation of the equivalence principle. This has motivated experimental searches for harmonic variations of the constants of Nature in a wide range of frequencies: with atomic clocks in the $10^{-10}-1$ Hz region \cite{Tilburg2015,Hees2016,Kennedy2020}, with a network of optical cavities in the $10^{-4}-10^{-1}$ Hz region \cite{Wcislo2018} and using atomic spectrocopy in the $10^5-10^8$ Hz region \cite{Antypas2019}. This has also given rise to various proposals for experiments to further extend the frequency range of searches for oscillations in the constants of Nature: using future atomic gravitational waves detector in the $10^{-3}-10^3$ Hz region \cite{Arvanitaki2018}, using future resonnant-mass detectors in the $10^4-10^6$ Hz region \cite{Arvanitaki2016}, using laser gravitational waves detector in the $10^{-2}-10^2$ Hz region \cite{StadnikLasInf2015,morisaki:2019aa,Grote2019a}, etc. In this paper, we propose a new type of experiment consisting of a three-arm Mach-Zender interferometer (see also the preprint \cite{Savalle_2019}) to search for harmonic variations in the fine-structure constant and in the mass of the electron in the partially unexplored $10^4-10^6$ Hz region. Furthermore, we present results from an experimental realisation developped at the Paris Observatory that provides the first constraints in the oscillations of the constants of Nature in the $10^4-10^5$ Hz frequency range and improves over previous results \cite{Antypas2019} in the $10^5-10^6$ Hz frequency range. Finally, we present an interpretation of these experimental results for different scenarios of DM.

\paragraph{\textbf{Ultra-light scalar field oscillation}}
The theory of ultra-light SF has been developed in, e.g. refs. \cite{Damour2010,StadnikDMalpha2015,Arvanitaki2014}. 
Within this framework, a scalar field $\varphi$ of mass $m_\varphi$ is linearly coupled to the standard model Lagrangian leading to space-time variations of any fundamental constant $X$ from the standard model.  The coupling parameters $d_X$ characterize the strength of the interaction between the SF and the various sectors of the standard model. More precisely, they parametrize the variation of the constants of Nature \cite{Damour2010} such that $X(\varphi) = X_0 \left(1 + d_X \varphi\right)$ where $X$ denotes any constant of Nature like e.g. the fine structure constant \{$\alpha$,$d_e$\}, the mass of fermions \{$m_j$,$d_{m_j}$\} ($j=$ electron, quarks) and the QCD mass scale \{$\Lambda_{QCD}$,$d_g$\} and the subscript $0$ refers to the value of the constant in absence of the SF.

In this paper, we will focus on low SF masses ($m_{\varphi}\ll 1$~eV) for which $\varphi$ behaves as a classical field. This SF admits oscillating solutions  $\varphi= \varphi_0 \cos(\omega_m t)$  (with $\omega_m$ the SF Compton-De Broglie frequency) \cite{Arvanitaki2014,StadnikDMalpha2015,Hees2016} which induces a temporal evolution of the constants of Nature. When the SF is interpreted as DM, its amplitude of  oscillation is directly related to the local DM density $\rho_\mathrm{DM}$ (0.4~GeV/cm$^3$  in the standard galactic halo DM model \cite{mcmillan:2011vn}). Variations of the fine structure constant and/or electron mass results in a variation of atomic transition frequencies and of the Bohr radius $a_0= \hbar/(m_e c \alpha)$, which in turn leads to variation of the frequency of atomic clocks and the length of solids. This has motivated several experimental searches for harmonic variations of the constants of Nature using various atomic experiments \cite{AtomicReview2017}.

In addition, searches for a Yukawa-like violation of the universality of free fall also provide constraints on the couplings between matter and the SF, see e.g. \cite{Wagner_2012,berge:2018aa,Hees2018}. Those constraints are independent of the identification of the SF as DM (and are therefore independent of the relative abundance and of the composition of DM).

\begin{figure}[h!]
\vspace{-0.5cm}
\includegraphics[width=0.45\textwidth]{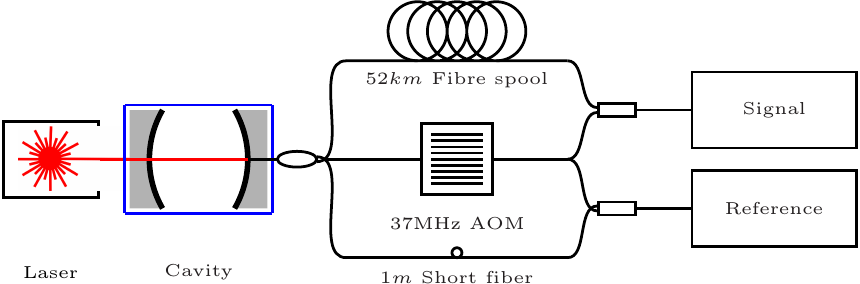}
\caption{Experimental setup. A $1542$nm laser source is locked to an ultra-stable cavity. The beam is then split between three arms and recombined to have access to the DM signal (long vs AOM arms) and the experimental reference (short vs AOM arms).}
\label{fig_Setup}
\end{figure}

\paragraph{\textbf{Experimental setup}}
Our experimental setup, dubbed the DAMNED (DArk Matter from Non Equal Delays) experiment is a three-arm Mach-Zender interferometer \cite{Savalle_2019} as shown in Fig. \ref{fig_Setup}. A $1542$nm laser source is stabilized on an ultra-stable  cavity \cite{Millo2009,Xie2017}, with a locking bandwidth of a few 100 kHz. The beam power is then unevenly distributed between the three arms. Most of the power is going through the long delay line that consists of a fibre spool (52 or 56 km) with a refractive index $n_0 \approx 1.5$. To perform a self-heterodyne detection, the laser frequency is shifted with the Acousto-optic modulator (AOM) located in the second arm (where $\nu_\mathrm{AOM}\approx37$~MHz). Finally, the last arm is a one meter fibre.

The beatnote between the AOM and the fibre spool arms provides the putative DM signal resulting from the beat between the (DM induced oscillating) cavity frequency at time $t$ and its frequency at time $t-T$ as seen through the fibre with delay $T$ (see next section for details). The reference beatnote between the AOM and the short fibre provides an indication of the experimental perturbations (noise and systematics) as the arm length is too short to see any cavity frequency oscillations. Both beatnotes are acquired simultaneously using a digital two channel phase-meter (Ettus X310) that provides the phase measurements (after demodulation of the 37~MHz signal) at a sampling rate of $500$kHz.

\paragraph{\textbf{Impact of an oscillating scalar field on the experiment}} \label{sec:theo_mod}
An oscillating scalar field will impact two parts of our experiment. First, it will lead to  oscillations of the cavity frequency $\omega(t)$ due to variations of its length induced by oscillations of the Bohr radius. Secondly, the fibre delay $T(t)$ will oscillate because of variations of its length and of its refraction index that are both induced by oscillations of the constants of Nature.

The laser is locked to an optical resonance of the cavity. The variation of the frequency of the light exiting the cavity $\delta \omega(t)$ is then proportional to its length variation and is given by
\begin{equation} \label{eq:omega(t)}
\frac{\delta \omega(t)}{\omega_0} = \epsilon_L \left(\mathcal{E}_c(1+\alpha)\cos(\omega_mt) + \mathcal{E}_s \beta \sin(\omega_mt)\right)\, ,
\end{equation}
where $\omega_0$ is the unperturbed cavity frequency, $\epsilon_L = \varphi_0 (d_e+d_{m_e})$ is the fractional length change due to oscillations of the Bohr radius. The coefficients $\mathcal{E}_c$ and $\mathcal{E}_s$ characterize the optical properties of the cavity.  For our high finesse cavity ($\mathcal{F} \approx 800000$ \cite{Millo2009}) and frequencies of interest ($f \in [10,200]$~kHz), we have $\mathcal{E}_c,\mathcal{E}_s \simeq 1$. On the other hand, the $\alpha$ and $\beta$ coefficients characterize the mechanical properties of the cavity. For our $\sim 0.1$~m ULE cavity the mechanical resonant frequencies are $\omega_n = 2 \pi n \, 27.6 $~kHz where $n$ is an integer ($n\geq 1$), and are therefore well within our frequency region of interest ($[10,200]$~kHz). Only odd resonances are excited due to the symmetry of the length change. At resonance ($\omega_m = \omega_1$), $\alpha \simeq 0$ and $\beta = 8Q_1/\pi^2$, with the quality factor of our ULE cavity $Q_1 = 6.1\times 10^4$ \cite{Millo2009,Numata2004,Zhang2013} which significantly enhances the signal searched for. Below resonance ($\omega_m \ll \omega_1$) both $\beta,\alpha \simeq 0$. A detailed derivation of the coefficients $\alpha,\beta,\mathcal{E}_c,\mathcal{E}_s$ is provided in the supplemental material \hypersetup{citecolor=blue}\cite{supplemental}\hypersetup{citecolor=green}. Similar analysis was also carried out in \cite{Arvanitaki2016,Grote2019a} giving similar results. Note that the mechanical resonant frequencies and Q-factors are determined from material dependent constants (see \hypersetup{citecolor=blue}\cite{supplemental}\hypersetup{citecolor=green}) and we conservatively assume a 5~\% uncertainty in those constants.

The fibre delay is given by $T(t) = L_f(t)n(t)/c$, where $L_f(t)$ and $n(t)$ are the fibre length and refractive index respectively, both of which oscillates due to the variations of the constants of Nature. Using the approach described in \cite{Braxmaier2001}, we find
\begin{equation} \label{eq:dT_T_1}
\frac{\delta T(t)}{T_{0}} =\frac{\omega_0}{n_0}\frac{\partial n}{\partial\omega}\left(\frac{\delta\omega(t)}{\omega_0} - \epsilon_n \cos(\omega_m t) \right) -\epsilon_L\cos(\omega_m t)\, , 
\end{equation}
where $\delta\omega(t)/\omega_0$ is the relative frequency variation at the entrance of the fibre, which in our case is given by Eq.~(\ref{eq:omega(t)}) and $\epsilon_n$ is the fractional refractive index change that are directly proportionnal to the amplitude of oscillations of the SF through
\begin{equation}
  \epsilon_n = \varphi_0 (2d_e + d_{m_e} + (d_{m_e}-d_g)/2 - 0.024(d_{m_q}-d_g))  \, .
\end{equation}
The dependence on $d_g$ and $d_{m_q}$ arises from the phonon frequencies in the fibre that determine its refractive index \cite{Braxmaier2001}. For the telecom fibres that we use the pre-factor of (\ref{eq:dT_T_1}) is typically $\approx 10^{-2}$.

Both the cavity frequency and fibre delay oscillations can be integrated to obtain the phase difference $\Delta \Phi (t)$ between the delayed and non-delayed signals :
\begin{equation}
\begin{aligned}
&\Delta \Phi (t) = \int_{t-T_0-\delta T(t)}^{t} \left(\omega_0 + \delta \omega(t')\right)\, \mathrm{d} t' \\
&= \omega_0 T_0 +2\frac{\omega_0}{\omega_m}\text{sin}\left(\frac{\omega_{m} T_0}{2}\right) \left[C_{\Delta \Phi} \cos\left(\omega_{m} t - \omega_{m} \frac{T_0}{2}\right) \right.\\
& \hspace{3.2 cm} +\left.S_{\Delta \Phi} \sin\left(\omega_{m} t - \omega_{m} \frac{T_0}{2}\right)\right] \, ,
\end{aligned}
\label{eq_final}
\end{equation}

where $C_{\Delta \Phi}$ and $S_{{\Delta \Phi}}$ are derived from (\ref{eq:omega(t)}) and (\ref{eq:dT_T_1}), to leading order :
\begin{equation} \label{eq:C-S}
\begin{aligned}
C_{\Delta \Phi} \simeq \epsilon_L\alpha - \epsilon_n \frac{\omega_0}{n_0}\frac{\partial n}{\partial\omega}\, ;\hspace{1.0 cm} S_{\Delta \Phi} \simeq \epsilon_L \beta  \, .
\end{aligned}
\end{equation}

Since (\ref{eq_final}) has extinctions for $\omega_m T_0  = n2\pi$, we use two different fibre lengths (thus two different delays $T_0$) to recover sensitivity over the whole desired frequency range. For the reference arm $T_0 \simeq 0$, and the signal of (\ref{eq_final}) vanishes, which allows its use to characterise systematic effects and identify false DM signals.

\paragraph{\textbf{Experimental results}} \label{sec:exp_res}
The parallel acquisitions of the signal and reference phase data lasted 12 days each for the two different fibre lengths ($52.64$ and $56.09$~km) at a $500$~kHz sampling rate. The total raw phase data ($\sim 4\times 2.1$~TB) requires digital pre-processing to compute Fourier transforms with a spectral resolution limited to $\sim3$~mHz \footnote{RAM limitation reduces the maximum amount of data duration that can be loaded to $\sim$268~s. The maximum spectral resolution is therefore $1/T_\mathrm{RAM} > 1/T_\mathrm{exp}$.}. 

\begin{figure}[h!]
\includegraphics[width=0.5\textwidth]{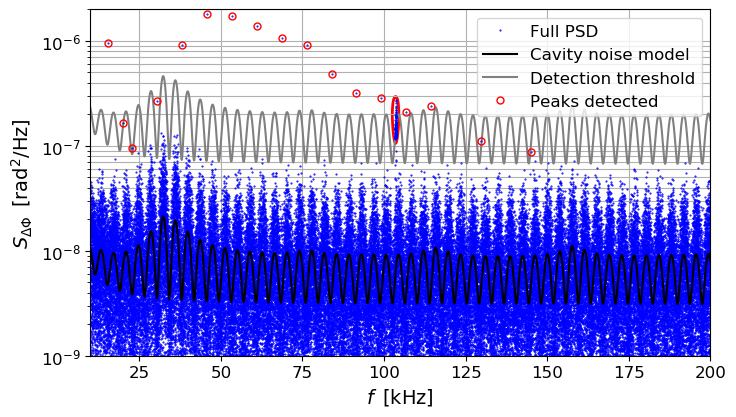}
\vspace{-1cm}
\caption{PSD of phase fluctuations $S_{\Delta \Phi}(f)$ of the signal for the 52~km fibre (in blue). The modelled cavity noise floor is shown in black, with the grey line indicating the 95\% detection threshold \cite{Scargle1982} which reveals several significant peaks (in red).}
\label{fig_BLUE_52}
\end{figure}

Figure \ref{fig_BLUE_52} shows the power spectral density (PSD) computed over the full 12 days duration of the experiment. Only the ``signal" branch for the $52$~km acquisition is shown here, but all results are similar for the $56$~km fibre, as well as for the ``reference" branch. One can see characteristic ``bumps" arising from the cavity noise seen through the transfer function of our unequal-length arm interferometer \hypersetup{citecolor=blue}\cite{supplemental}\hypersetup{citecolor=green}. Our experimental sensitivity is limited below $10$~kHz by the acoustic and thermal noise of the long signal fibre and above $200$~kHz by the bandwidth of the phase-locked-loop that locks the laser to the cavity (which we optimised in our experiment to $\approx 500$~kHz). 
In between, the experiment is limited by the cavity noise floor which was stationary during the full duration of both acquisitions. The modelled cavity noise is an averaged PSD over $268$~s data subsets and is required later for our statistical analysis (see \hypersetup{citecolor=blue}\cite{supplemental}\hypersetup{citecolor=green} for details).

\paragraph{\textbf{Systematic effects}} 
In order to identify a potential DM signal, we investigate any signal (peak in the PSD) emerging from the noise. For this, we use the method presented in \cite{Scargle1982} to define a detection level above which any peak can be considered as a real signal with a false detection rate lower than 5\%. As one can see  in Fig. \ref{fig_BLUE_52}, the detection threshold (grey line), is exceeded by many peaks (red dots). At closer inspection (see \hypersetup{citecolor=blue}\cite{supplemental}\hypersetup{citecolor=green} for details) all of these peaks turn out to be systematic effects that are either present only in the cavity used for the experiment and not in the other cavities available in the laboratory, and/or are correlated to temperature changes in the laboratory, or are also present in the ``reference" branch which is insensitive to coupling with DM. Additionally, they have a spectral shape and width incompatible with the signal in the DM models we consider. Therefore, we report no detection of ultra-light DM in the frequency range $\left[10,200\right]\,$~kHz. However, that conclusion does not apply at the frequencies of the systematic peaks, which might mask a putative DM signal, and we thus exclude those frequency regions from our results, as summarised in tab. \ref{tab_peaks}. All the peak positions are drifting relatively to their mean value $\langle f \rangle$ by a factor $r_f$. With the peak widths $\sigma_f$, we define a conservative exclusion interval $[\langle f \rangle \times (1-r_f) - 3 \sigma_f,\langle f \rangle \times (1+r_f) + 3 \sigma_f]$.

\begin{table}[h]
\begin{tabular}{l c c c l r}
Origin & $\langle f \rangle~/$Hz & $r_f$ & $\sigma_f$/Hz & $[f_{min}/$Hz, & $f_{max}/$Hz] \\
 \hline  
 & 			$26178$ & $10^{-4}$ & $1$ & $[26172.382,$ & $26183.618]$\\
Unknown &  	$50069$ & $10^{-4}$ & $1$ & $[50060.993,$ & $50077.007]$\\
 & 			$59364$ & $10^{-4}$ & $1$ & $[59355.064,$ & $59372.936]$\\
 \hline  
Piezo &  	$101684$ & $10^{-3}$ & $3$ & $[101573.316,$ & $101794.684]$\\
 & 			$103525$ & $10^{-3}$ & $3$ & $[103412.975,$ & $103638.025]$\\
 \hline  
Ettus &  	\multicolumn{5}{c}{Multiples of $7629.395$~Hz, $\sigma_f = 3$~mHz}\\
\end{tabular}
\caption{Excluded frequency regions due to systematic effects.}
\label{tab_peaks}
\end{table}

\paragraph{\textbf{Constraints on ultra-light Dark Matter}} The first DM model for which we interpret our measurements is a standard model where all the DM density is assumed to be uniformly distributed in the solar System and is carried by the scalar field~\cite{Arvanitaki2014,StadnikDMalpha2015,Hees2018}. In standard models of galaxy formation, galactic DM must be virialized \cite{Freese2013,Pillepich2014} and has a velocity distribution $f_\mathrm{DM}(v)$ with a characteristic width $\sigma_v \sim 10^{-3}c$ \cite{Derevianko2016,Foster2018,evans:2019aa}. The Compton frequency of the SF, $\omega_m \simeq m_{\varphi} c^2/\hbar  \left(1 +  v^2/(2c^2)\right)$, is broadened because of the DM velocity distribution. This broadening introduces a coherence time $\tau_c = (\omega_m \sigma^2_v/c^2)^{-1}\sim 10^{6} \omega_m^{-1}$ \cite{DereviankoVULF2016}. The DM distribution therefore implies that the scalar field has a stochastic component from the sum of all the SF allowed by the velocity distribution. The effective field takes the following form \cite{Foster2018,Centers2019} : 
	\begin{equation} \label{eq:field_s} 
		\varphi(t)= \frac{ \sqrt{ \frac{4\pi \rho_\mathrm{DM} }{c^2}}}{\frac{m_\varphi c^2}{\hbar}} \sum_{j=1}^{N_j} \alpha_j \sqrt{f_\mathrm{DM}(f_j)\Delta f} \cos \left[2\pi f_j t+\phi_j\right]\, ,
	\end{equation}
where $\alpha_j$ are stochastic amplitudes following a Rayleigh distribution \cite{Foster2018,Centers2019}, $\phi_j$ are random phases following a uniform distribution and $f_\mathrm{DM}(f)$ is the DM velocity distribution expressed in the frequency domain (see \hypersetup{citecolor=blue}\cite{supplemental}\hypersetup{citecolor=green} and \cite{Derevianko2016}). $N_j$ defines the number of points used to discretize the DM frequency distribution curve ($N_j \Delta f \geq 1/\tau_c$, where $\Delta f$ is the frequency resolution of the data). When the experimental duration $T_\mathrm{exp}$ is longer than $\tau_c$, this stochastic broadening needs to be taken into account in the data analysis \cite{Derevianko2016,Foster2018} and actually provides a useful handle on identifying the signal due to its peculiar spectral shape. Even when $T_\mathrm{exp} \leq \tau_c$, the stochastic nature of the signal needs to be taken into account as in general it leads to reduced sensitivity by up to three orders of magnitude because of the possibility of the instantaneous local oscillation amplitude being smaller than the average value which is related to $\rho_\mathrm{DM}$ \cite{Centers2019}.

\begin{figure*}[t]
\clearpage\includegraphics[width=\textwidth]{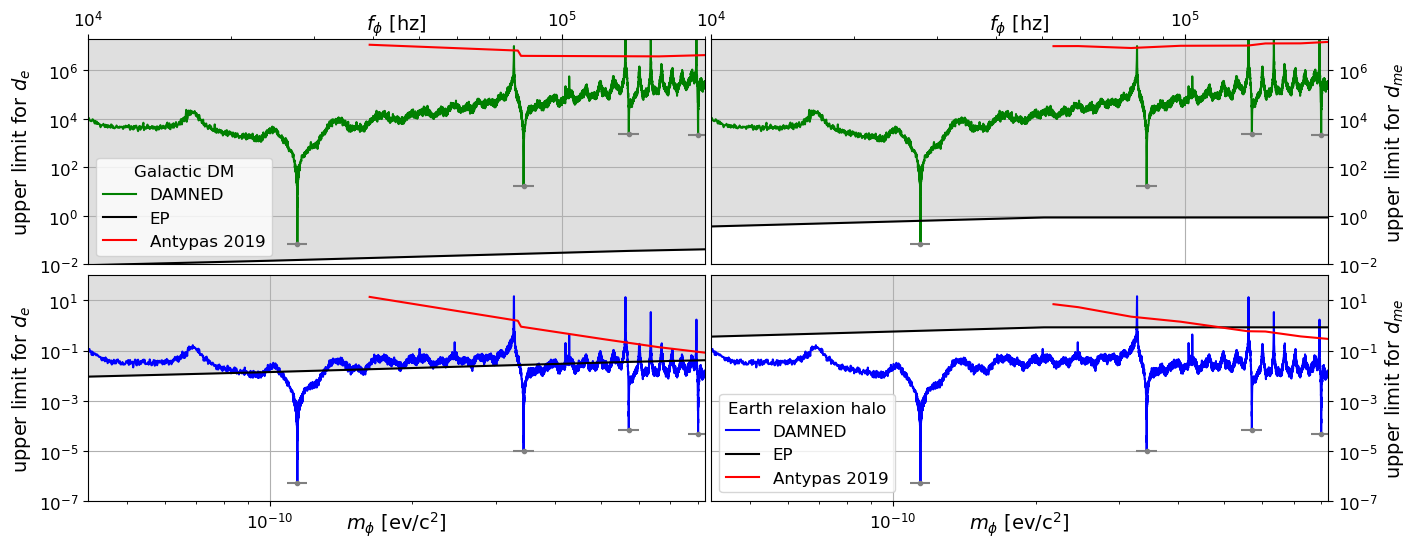}
\caption{95\% confidence upper limits on $d_e$ and $d_{m_e}$ in the usual galactic DM model (top) and in the Earth relaxion halo one (bottom). Sensitivity peaks are at the mechanical resonance frequencies of the cavity. The solid black line corresponds to the constraints set by the E\"ot-Wash torsion balance experiments \cite{Schlamminger2008,Wagner_2012,Hees2018} while the red line corresponds to a more recent experiment  \cite{Antypas2019}. Narrow frequency bands excluded from these constraints because of systematic effects are given in Tab. \ref{tab_peaks}. Note that the frequency and amplitude of the resonance peaks may differ from those shown by $\sim$5~\% because of the uncertainty in $\omega_n$ and $Q_n$.}
\label{fig_limits}
\end{figure*}

In order to constrain the DM model, the coupling constants must be extracted from the coefficients $C_{\Delta \Phi}$ and $S_{\Delta \Phi}$ available in the Fourier transform of our data. The stochastic nature of the signal (\ref{eq:field_s}) requires the adjustment of the following parameters: the coupling constants $d_X$, $N_j$ amplitudes $\alpha_j$ and $N_j$ phases $\phi_j$, where $N_j$ is chosen to sufficiently sample the DM frequency distribution $f_\mathrm{DM}$. The a priori knowledge of the probability distribution of amplitudes (Rayleigh distribution) and phases (uniform distribution) favours the use of a Bayesian approach. Working in the frequency domain the corresponding posterior distributions can be analytically marginalised over the $N_j$ amplitudes $\alpha_j$ and phases $\phi_j$, which makes the problem numerically solvable (see \cite{Foster2018,Derevianko2016,Centers2019} and \hypersetup{citecolor=blue}\cite{supplemental}\hypersetup{citecolor=green}). The result is a posterior probability distribution for the coupling constants $d_X$ for each DM mass, providing the corresponding 95\% upper limit.  To simplify, we concentrate on $d_e$ and $d_{m_e}$ and assume that only one of them is non-zero in turn, a so called ``maximum reach approach". We use our acquisitions with two different fibre lengths and combine both likelihoods to infer a unique upper limit at 95\% confidence. These upper limits for the galactic DM model (where we assumed that the scalar field is made from 100\% of the the DM energy density) are presented in the top part of Fig. \ref{fig_limits}. The constraints show large ``peaks" at the resonant frequencies ($n=1,3,5,7$) of our cavity, and at frequencies where the combination of two different fibre lengths does not fully solve the loss of sensitivity due to the $\sin(\omega_mT_0/2)$ term in (\ref{eq_final}). In between the peaks the constraints come from a combination of the length and index changes of the cavity and the fibres. For this specific theoretical scenario, our experiment exceeds best existing constraints on $d_{m_e}$ from torsion balance experiments \cite{Schlamminger2008,Wagner_2012,Hees2018} by about an order of magnitude only over a narrow-frequency band around the cavity resonance, but broadly improves on the recent experiment reported in \cite{Antypas2019}, by up to three orders of magnitude\footnote{Note that in \cite{Antypas2019} the authors do not take the factor $\sim$10 sensitivity loss pointed out in \cite{Centers2019} into account, contrary to our work (see \hypersetup{citecolor=blue}\cite{supplemental}\hypersetup{citecolor=green}).}.

\paragraph{\textbf{Constraints on the relaxion halo model}} 
The second theoretical model for which we interpret our experimental results is called relaxion halo model \cite{Banerjee2020}. In this scenario, DM forms a relaxion halo around the Earth \cite{Kolb1993,Levkov2018,Braaten2019,Vaquero2019,BarOr2019} leading to a local overdensity with respect to the galactic DM density that depends on $m_\varphi$ and can reach $\rho_\mathrm{RH}/\rho_\mathrm{DM} \leq 10^{16}$ in the range of $m_\varphi$ considered here \cite{Banerjee2020}. Additionally the velocity distribution, and therefore the coherence time is modified, leading to $\tau_c \sim 10^{20}\omega_m^{-1} (2\pi\,\mathrm{Hz}/\omega_m)^2$. Both of these modifications have to be taken into account in the data analysis. First experimental searches in this model were reported in \cite{Aharony2019,Antypas2019}. We present the constraint on the coupling parameters obtained in this scenario in the bottom of Fig.~\ref{fig_limits}. In this model, our experiment improves on best existing constraints for almost all of the probed DM masses. That improvement reaches 5 orders of magnitude for $d_{e}$ and 6 orders of magnitude for $d_{m_e}$ at the mechanical resonances.

The underlying reason for the difference in sensitivity in the two models comes from the fact that experiments like ours or \cite{Antypas2019} depend on the local DM density while torsion balance experiments search for a Yukawa interaction between the Earth and the test masses, which is independent of the identification of the SF as DM \cite{Hees2018} and are thus independent of the local DM density and composition.

In all cases our experiment improves on the recent experiment reported in \cite{Antypas2019}, which directly probes the same DM models as ours, by typically 2-3 orders of magnitude over the DM mass region where the two overlap.

\paragraph{\textbf{Conclusion}} In this letter, we propose a new experiment to search for harmonic variation of the constants of Nature at high frequencies. In addition, we present results from the DAMNED experiment developed at the Paris Observatory. This experiment has not revealed any sign of scalar DM for masses in the [$4.1\times 10^{-11}$, $8.3\times 10^{-10}$]~eV region, but we have improved existing bounds on the DM-SM coupling constants by amounts depending on the considered mass and DM distribution model.

Our main limitation is the cavity noise, and we plan to improve on the results presented here over the next years, and also test other models (e.g. axion couplings), using similar set-ups but with an improved optical cavity currently under construction.

\begin{acknowledgments}
\paragraph{\textbf{Acknowledgments}} Helpful discussions with Andrei Derevianko and Yevgeny Stadnik are gratefully acknowledged.
\end{acknowledgments}

\onecolumngrid

\clearpage
\newpage
\nocite{Lee2016,Vinet2010,Canuel2017,Banerjee2019}
\bibliography{Bib_DAMNED_PRL}

\begin{thebibliography}{53}%
\makeatletter
\providecommand \@ifxundefined [1]{%
 \@ifx{#1\undefined}
}%
\providecommand \@ifnum [1]{%
 \ifnum #1\expandafter \@firstoftwo
 \else \expandafter \@secondoftwo
 \fi
}%
\providecommand \@ifx [1]{%
 \ifx #1\expandafter \@firstoftwo
 \else \expandafter \@secondoftwo
 \fi
}%
\providecommand \natexlab [1]{#1}%
\providecommand \enquote  [1]{``#1''}%
\providecommand \bibnamefont  [1]{#1}%
\providecommand \bibfnamefont [1]{#1}%
\providecommand \citenamefont [1]{#1}%
\providecommand \href@noop [0]{\@secondoftwo}%
\providecommand \href [0]{\begingroup \@sanitize@url \@href}%
\providecommand \@href[1]{\@@startlink{#1}\@@href}%
\providecommand \@@href[1]{\endgroup#1\@@endlink}%
\providecommand \@sanitize@url [0]{\catcode `\\12\catcode `\$12\catcode
  `\&12\catcode `\#12\catcode `\^12\catcode `\_12\catcode `\%12\relax}%
\providecommand \@@startlink[1]{}%
\providecommand \@@endlink[0]{}%
\providecommand \url  [0]{\begingroup\@sanitize@url \@url }%
\providecommand \@url [1]{\endgroup\@href {#1}{\urlprefix }}%
\providecommand \urlprefix  [0]{URL }%
\providecommand \Eprint [0]{\href }%
\providecommand \doibase [0]{http://dx.doi.org/}%
\providecommand \selectlanguage [0]{\@gobble}%
\providecommand \bibinfo  [0]{\@secondoftwo}%
\providecommand \bibfield  [0]{\@secondoftwo}%
\providecommand \translation [1]{[#1]}%
\providecommand \BibitemOpen [0]{}%
\providecommand \bibitemStop [0]{}%
\providecommand \bibitemNoStop [0]{.\EOS\space}%
\providecommand \EOS [0]{\spacefactor3000\relax}%
\providecommand \BibitemShut  [1]{\csname bibitem#1\endcsname}%
\let\auto@bib@innerbib\@empty
\bibitem [{\citenamefont {Bertone}\ and\ \citenamefont
  {Tait}(2018)}]{Bertone2018}%
  \BibitemOpen
  \bibfield  {author} {\bibinfo {author} {\bibfnamefont {G.}~\bibnamefont
  {Bertone}}\ and\ \bibinfo {author} {\bibfnamefont {T.~M.~P.}\ \bibnamefont
  {Tait}},\ }\href {\doibase 10.1038/s41586-018-0542-z} {\bibfield  {journal}
  {\bibinfo  {journal} {Nature}\ }\textbf {\bibinfo {volume} {562}},\ \bibinfo
  {pages} {51} (\bibinfo {year} {2018})},\ \Eprint
  {http://arxiv.org/abs/1810.01668} {arXiv:1810.01668} \BibitemShut {NoStop}%
\bibitem [{\citenamefont {Safronova}\ \emph {et~al.}(2018)\citenamefont
  {Safronova}, \citenamefont {Budker}, \citenamefont {DeMille}, \citenamefont
  {Kimball}, \citenamefont {Derevianko},\ and\ \citenamefont
  {Clark}}]{AtomicReview2017}%
  \BibitemOpen
  \bibfield  {author} {\bibinfo {author} {\bibfnamefont {M.~S.}\ \bibnamefont
  {Safronova}}, \bibinfo {author} {\bibfnamefont {D.}~\bibnamefont {Budker}},
  \bibinfo {author} {\bibfnamefont {D.}~\bibnamefont {DeMille}}, \bibinfo
  {author} {\bibfnamefont {D.~F.~J.}\ \bibnamefont {Kimball}}, \bibinfo
  {author} {\bibfnamefont {A.}~\bibnamefont {Derevianko}}, \ and\ \bibinfo
  {author} {\bibfnamefont {C.~W.}\ \bibnamefont {Clark}},\ }\href {\doibase
  10.1103/RevModPhys.90.025008} {\bibfield  {journal} {\bibinfo  {journal}
  {Rev. Mod. Phys.}\ }\textbf {\bibinfo {volume} {90}},\ \bibinfo {pages}
  {025008} (\bibinfo {year} {2018})}\BibitemShut {NoStop}%
\bibitem [{\citenamefont {Derevianko}\ and\ \citenamefont
  {Pospelov}(2014)}]{DereviankoDM2014}%
  \BibitemOpen
  \bibfield  {author} {\bibinfo {author} {\bibfnamefont {A.}~\bibnamefont
  {Derevianko}}\ and\ \bibinfo {author} {\bibfnamefont {M.}~\bibnamefont
  {Pospelov}},\ }\href {\doibase 10.1038/nphys3137} {\bibfield  {journal}
  {\bibinfo  {journal} {Nat. Phys.}\ }\textbf {\bibinfo {volume} {10}},\
  \bibinfo {pages} {933} (\bibinfo {year} {2014})}\BibitemShut {NoStop}%
\bibitem [{\citenamefont {Arvanitaki}\ \emph {et~al.}(2015)\citenamefont
  {Arvanitaki}, \citenamefont {Huang},\ and\ \citenamefont {{Van
  Tilburg}}}]{Arvanitaki2014}%
  \BibitemOpen
  \bibfield  {author} {\bibinfo {author} {\bibfnamefont {A.}~\bibnamefont
  {Arvanitaki}}, \bibinfo {author} {\bibfnamefont {J.}~\bibnamefont {Huang}}, \
  and\ \bibinfo {author} {\bibfnamefont {K.}~\bibnamefont {{Van Tilburg}}},\
  }\href {\doibase 10.1103/PhysRevD.91.015015} {\bibfield  {journal} {\bibinfo
  {journal} {Phys. Rev. D}\ }\textbf {\bibinfo {volume} {91}},\ \bibinfo
  {pages} {015015} (\bibinfo {year} {2015})}\BibitemShut {NoStop}%
\bibitem [{\citenamefont {Stadnik}\ and\ \citenamefont
  {Flambaum}(2015)}]{StadnikDMalpha2015}%
  \BibitemOpen
  \bibfield  {author} {\bibinfo {author} {\bibfnamefont {Y.~V.}\ \bibnamefont
  {Stadnik}}\ and\ \bibinfo {author} {\bibfnamefont {V.~V.}\ \bibnamefont
  {Flambaum}},\ }\href {\doibase 10.1103/PhysRevLett.115.201301} {\bibfield
  {journal} {\bibinfo  {journal} {Phys. Rev. Lett.}\ }\textbf {\bibinfo
  {volume} {115}},\ \bibinfo {pages} {201301} (\bibinfo {year}
  {2015})}\BibitemShut {NoStop}%
\bibitem [{\citenamefont {{Van Tilburg}}\ \emph {et~al.}(2015)\citenamefont
  {{Van Tilburg}}, \citenamefont {Leefer}, \citenamefont {Bougas},\ and\
  \citenamefont {Budker}}]{Tilburg2015}%
  \BibitemOpen
  \bibfield  {author} {\bibinfo {author} {\bibfnamefont {K.}~\bibnamefont {{Van
  Tilburg}}}, \bibinfo {author} {\bibfnamefont {N.}~\bibnamefont {Leefer}},
  \bibinfo {author} {\bibfnamefont {L.}~\bibnamefont {Bougas}}, \ and\ \bibinfo
  {author} {\bibfnamefont {D.}~\bibnamefont {Budker}},\ }\href {\doibase
  10.1103/PhysRevLett.115.011802} {\bibfield  {journal} {\bibinfo  {journal}
  {Phys. Rev. Lett.}\ }\textbf {\bibinfo {volume} {115}},\ \bibinfo {pages}
  {011802} (\bibinfo {year} {2015})}\BibitemShut {NoStop}%
\bibitem [{\citenamefont {Hees}\ \emph {et~al.}(2016)\citenamefont {Hees},
  \citenamefont {Gu{\'{e}}na}, \citenamefont {Abgrall}, \citenamefont {Bize},\
  and\ \citenamefont {Wolf}}]{Hees2016}%
  \BibitemOpen
  \bibfield  {author} {\bibinfo {author} {\bibfnamefont {A.}~\bibnamefont
  {Hees}}, \bibinfo {author} {\bibfnamefont {J.}~\bibnamefont {Gu{\'{e}}na}},
  \bibinfo {author} {\bibfnamefont {M.}~\bibnamefont {Abgrall}}, \bibinfo
  {author} {\bibfnamefont {S.}~\bibnamefont {Bize}}, \ and\ \bibinfo {author}
  {\bibfnamefont {P.}~\bibnamefont {Wolf}},\ }\href {\doibase
  10.1103/PhysRevLett.117.061301} {\bibfield  {journal} {\bibinfo  {journal}
  {Phys. Rev. Lett.}\ }\textbf {\bibinfo {volume} {117}},\ \bibinfo {pages}
  {061301} (\bibinfo {year} {2016})}\BibitemShut {NoStop}%
\bibitem [{\citenamefont {Wcis{\l}o}\ \emph {et~al.}(2016)\citenamefont
  {Wcis{\l}o}, \citenamefont {Morzy{\'{n}}ski}, \citenamefont {Bober},
  \citenamefont {Cygan}, \citenamefont {Lisak}, \citenamefont {Ciury{\l}o},\
  and\ \citenamefont {Zawada}}]{Wcislo2016}%
  \BibitemOpen
  \bibfield  {author} {\bibinfo {author} {\bibfnamefont {P.}~\bibnamefont
  {Wcis{\l}o}}, \bibinfo {author} {\bibfnamefont {P.}~\bibnamefont
  {Morzy{\'{n}}ski}}, \bibinfo {author} {\bibfnamefont {M.}~\bibnamefont
  {Bober}}, \bibinfo {author} {\bibfnamefont {A.}~\bibnamefont {Cygan}},
  \bibinfo {author} {\bibfnamefont {D.}~\bibnamefont {Lisak}}, \bibinfo
  {author} {\bibfnamefont {R.}~\bibnamefont {Ciury{\l}o}}, \ and\ \bibinfo
  {author} {\bibfnamefont {M.}~\bibnamefont {Zawada}},\ }\href {\doibase
  10.1038/s41550-016-0009} {\bibfield  {journal} {\bibinfo  {journal} {Nat.
  Astron.}\ }\textbf {\bibinfo {volume} {1}},\ \bibinfo {pages} {0009}
  (\bibinfo {year} {2016})}\BibitemShut {NoStop}%
\bibitem [{\citenamefont {Roberts}\ \emph {et~al.}(2017)\citenamefont
  {Roberts}, \citenamefont {Blewitt}, \citenamefont {Dailey}, \citenamefont
  {Murphy}, \citenamefont {Pospelov}, \citenamefont {Rollings}, \citenamefont
  {Sherman}, \citenamefont {Williams},\ and\ \citenamefont
  {Derevianko}}]{GPSDM2017}%
  \BibitemOpen
  \bibfield  {author} {\bibinfo {author} {\bibfnamefont {B.~M.}\ \bibnamefont
  {Roberts}}, \bibinfo {author} {\bibfnamefont {G.}~\bibnamefont {Blewitt}},
  \bibinfo {author} {\bibfnamefont {C.}~\bibnamefont {Dailey}}, \bibinfo
  {author} {\bibfnamefont {M.}~\bibnamefont {Murphy}}, \bibinfo {author}
  {\bibfnamefont {M.}~\bibnamefont {Pospelov}}, \bibinfo {author}
  {\bibfnamefont {A.}~\bibnamefont {Rollings}}, \bibinfo {author}
  {\bibfnamefont {J.}~\bibnamefont {Sherman}}, \bibinfo {author} {\bibfnamefont
  {W.}~\bibnamefont {Williams}}, \ and\ \bibinfo {author} {\bibfnamefont
  {A.}~\bibnamefont {Derevianko}},\ }\href {\doibase
  10.1038/s41467-017-01440-4} {\bibfield  {journal} {\bibinfo  {journal} {Nat.
  Commun.}\ }\textbf {\bibinfo {volume} {8}},\ \bibinfo {pages} {1195}
  (\bibinfo {year} {2017})}\BibitemShut {NoStop}%
\bibitem [{\citenamefont {Hees}\ \emph {et~al.}(2018)\citenamefont {Hees},
  \citenamefont {Minazzoli}, \citenamefont {Savalle}, \citenamefont {Stadnik},\
  and\ \citenamefont {Wolf}}]{Hees2018}%
  \BibitemOpen
  \bibfield  {author} {\bibinfo {author} {\bibfnamefont {A.}~\bibnamefont
  {Hees}}, \bibinfo {author} {\bibfnamefont {O.}~\bibnamefont {Minazzoli}},
  \bibinfo {author} {\bibfnamefont {E.}~\bibnamefont {Savalle}}, \bibinfo
  {author} {\bibfnamefont {Y.~V.}\ \bibnamefont {Stadnik}}, \ and\ \bibinfo
  {author} {\bibfnamefont {P.}~\bibnamefont {Wolf}},\ }\href {\doibase
  10.1103/PhysRevD.98.064051} {\bibfield  {journal} {\bibinfo  {journal} {Phys.
  Rev. D}\ }\textbf {\bibinfo {volume} {98}},\ \bibinfo {pages} {064051}
  (\bibinfo {year} {2018})}\BibitemShut {NoStop}%
\bibitem [{\citenamefont {Wcis{\l}o}\ \emph {et~al.}(2018)\citenamefont
  {Wcis{\l}o}, \citenamefont {Ablewski}, \citenamefont {Beloy}, \citenamefont
  {Bilicki}, \citenamefont {Bober}, \citenamefont {Brown}, \citenamefont
  {Fasano}, \citenamefont {Ciury{\l}o}, \citenamefont {Hachisu}, \citenamefont
  {Ido}, \citenamefont {Lodewyck}, \citenamefont {Ludlow}, \citenamefont
  {McGrew}, \citenamefont {Morzy{\'n}ski}, \citenamefont {Nicolodi},
  \citenamefont {Schioppo}, \citenamefont {Sekido}, \citenamefont {Le~Targat},
  \citenamefont {Wolf}, \citenamefont {Zhang}, \citenamefont {Zjawin},\ and\
  \citenamefont {Zawada}}]{Wcislo2018}%
  \BibitemOpen
  \bibfield  {author} {\bibinfo {author} {\bibfnamefont {P.}~\bibnamefont
  {Wcis{\l}o}}, \bibinfo {author} {\bibfnamefont {P.}~\bibnamefont {Ablewski}},
  \bibinfo {author} {\bibfnamefont {K.}~\bibnamefont {Beloy}}, \bibinfo
  {author} {\bibfnamefont {S.}~\bibnamefont {Bilicki}}, \bibinfo {author}
  {\bibfnamefont {M.}~\bibnamefont {Bober}}, \bibinfo {author} {\bibfnamefont
  {R.}~\bibnamefont {Brown}}, \bibinfo {author} {\bibfnamefont
  {R.}~\bibnamefont {Fasano}}, \bibinfo {author} {\bibfnamefont
  {R.}~\bibnamefont {Ciury{\l}o}}, \bibinfo {author} {\bibfnamefont
  {H.}~\bibnamefont {Hachisu}}, \bibinfo {author} {\bibfnamefont
  {T.}~\bibnamefont {Ido}}, \bibinfo {author} {\bibfnamefont {J.}~\bibnamefont
  {Lodewyck}}, \bibinfo {author} {\bibfnamefont {A.}~\bibnamefont {Ludlow}},
  \bibinfo {author} {\bibfnamefont {W.}~\bibnamefont {McGrew}}, \bibinfo
  {author} {\bibfnamefont {P.}~\bibnamefont {Morzy{\'n}ski}}, \bibinfo {author}
  {\bibfnamefont {D.}~\bibnamefont {Nicolodi}}, \bibinfo {author}
  {\bibfnamefont {M.}~\bibnamefont {Schioppo}}, \bibinfo {author}
  {\bibfnamefont {M.}~\bibnamefont {Sekido}}, \bibinfo {author} {\bibfnamefont
  {R.}~\bibnamefont {Le~Targat}}, \bibinfo {author} {\bibfnamefont
  {P.}~\bibnamefont {Wolf}}, \bibinfo {author} {\bibfnamefont {X.}~\bibnamefont
  {Zhang}}, \bibinfo {author} {\bibfnamefont {B.}~\bibnamefont {Zjawin}}, \
  and\ \bibinfo {author} {\bibfnamefont {M.}~\bibnamefont {Zawada}},\
  }\href@noop {} {\bibfield  {journal} {\bibinfo  {journal} {Science Advances}\
  }\textbf {\bibinfo {volume} {4}},\ \bibinfo {pages} {eaau4869} (\bibinfo
  {year} {2018})}\BibitemShut {NoStop}%
\bibitem [{\citenamefont {Roberts}\ and\ \citenamefont
  {Derevianko}(2018)}]{RobertsAsymm2018}%
  \BibitemOpen
  \bibfield  {author} {\bibinfo {author} {\bibfnamefont {B.~M.}\ \bibnamefont
  {Roberts}}\ and\ \bibinfo {author} {\bibfnamefont {A.}~\bibnamefont
  {Derevianko}},\ }\href@noop {} {\enquote {\bibinfo {title} {Precision
  measurement noise asymmetry and its annual modulation as a dark matter
  signature},}\ } (\bibinfo {year} {2018}),\ \Eprint
  {http://arxiv.org/abs/1803.00617} {arXiv:1803.00617 [physics.atom-ph]}
  \BibitemShut {NoStop}%
\bibitem [{\citenamefont {Roberts}\ \emph {et~al.}(2018)\citenamefont
  {Roberts}, \citenamefont {Blewitt}, \citenamefont {Dailey},\ and\
  \citenamefont {Derevianko}}]{GPSDM2018}%
  \BibitemOpen
  \bibfield  {author} {\bibinfo {author} {\bibfnamefont {B.~M.}\ \bibnamefont
  {Roberts}}, \bibinfo {author} {\bibfnamefont {G.}~\bibnamefont {Blewitt}},
  \bibinfo {author} {\bibfnamefont {C.}~\bibnamefont {Dailey}}, \ and\ \bibinfo
  {author} {\bibfnamefont {A.}~\bibnamefont {Derevianko}},\ }\href {\doibase
  10.1103/PhysRevD.97.083009} {\bibfield  {journal} {\bibinfo  {journal} {Phys.
  Rev. D}\ }\textbf {\bibinfo {volume} {97}},\ \bibinfo {pages} {083009}
  (\bibinfo {year} {2018})}\BibitemShut {NoStop}%
\bibitem [{\citenamefont {Alonso}\ \emph {et~al.}(2019)\citenamefont {Alonso},
  \citenamefont {Blas},\ and\ \citenamefont {Wolf}}]{Alonso2019}%
  \BibitemOpen
  \bibfield  {author} {\bibinfo {author} {\bibfnamefont {R.}~\bibnamefont
  {Alonso}}, \bibinfo {author} {\bibfnamefont {D.}~\bibnamefont {Blas}}, \ and\
  \bibinfo {author} {\bibfnamefont {P.}~\bibnamefont {Wolf}},\ }\href {\doibase
  10.1007/JHEP07(2019)069} {\bibfield  {journal} {\bibinfo  {journal} {Journal
  of High Energy Physics}\ }\textbf {\bibinfo {volume} {2019}},\ \bibinfo
  {pages} {69} (\bibinfo {year} {2019})}\BibitemShut {NoStop}%
\bibitem [{\citenamefont {Wolf}\ \emph {et~al.}(2019)\citenamefont {Wolf},
  \citenamefont {Alonso},\ and\ \citenamefont {Blas}}]{Wolf2019}%
  \BibitemOpen
  \bibfield  {author} {\bibinfo {author} {\bibfnamefont {P.}~\bibnamefont
  {Wolf}}, \bibinfo {author} {\bibfnamefont {R.}~\bibnamefont {Alonso}}, \ and\
  \bibinfo {author} {\bibfnamefont {D.}~\bibnamefont {Blas}},\ }\href {\doibase
  10.1103/PhysRevD.99.095019} {\bibfield  {journal} {\bibinfo  {journal} {Phys.
  Rev. D}\ }\textbf {\bibinfo {volume} {99}},\ \bibinfo {pages} {095019}
  (\bibinfo {year} {2019})}\BibitemShut {NoStop}%
\bibitem [{\citenamefont {Kennedy}\ \emph {et~al.}(2020)\citenamefont
  {Kennedy}, \citenamefont {Oelker}, \citenamefont {Robinson}, \citenamefont
  {Bothwell}, \citenamefont {Kedar}, \citenamefont {Milner}, \citenamefont
  {Marti}, \citenamefont {Derevianko},\ and\ \citenamefont {Ye}}]{Kennedy2020}%
  \BibitemOpen
  \bibfield  {author} {\bibinfo {author} {\bibfnamefont {C.~J.}\ \bibnamefont
  {Kennedy}}, \bibinfo {author} {\bibfnamefont {E.}~\bibnamefont {Oelker}},
  \bibinfo {author} {\bibfnamefont {J.~M.}\ \bibnamefont {Robinson}}, \bibinfo
  {author} {\bibfnamefont {T.}~\bibnamefont {Bothwell}}, \bibinfo {author}
  {\bibfnamefont {D.}~\bibnamefont {Kedar}}, \bibinfo {author} {\bibfnamefont
  {W.~R.}\ \bibnamefont {Milner}}, \bibinfo {author} {\bibfnamefont {G.~E.}\
  \bibnamefont {Marti}}, \bibinfo {author} {\bibfnamefont {A.}~\bibnamefont
  {Derevianko}}, \ and\ \bibinfo {author} {\bibfnamefont {J.}~\bibnamefont
  {Ye}},\ }\href {\doibase 10.1103/PhysRevLett.125.201302} {\bibfield
  {journal} {\bibinfo  {journal} {Phys. Rev. Lett.}\ }\textbf {\bibinfo
  {volume} {125}},\ \bibinfo {pages} {201302} (\bibinfo {year}
  {2020})}\BibitemShut {NoStop}%
\bibitem [{\citenamefont {Antypas}\ \emph {et~al.}(2019)\citenamefont
  {Antypas}, \citenamefont {Tretiak}, \citenamefont {Garcon}, \citenamefont
  {Ozeri}, \citenamefont {Perez},\ and\ \citenamefont {Budker}}]{Antypas2019}%
  \BibitemOpen
  \bibfield  {author} {\bibinfo {author} {\bibfnamefont {D.}~\bibnamefont
  {Antypas}}, \bibinfo {author} {\bibfnamefont {O.}~\bibnamefont {Tretiak}},
  \bibinfo {author} {\bibfnamefont {A.}~\bibnamefont {Garcon}}, \bibinfo
  {author} {\bibfnamefont {R.}~\bibnamefont {Ozeri}}, \bibinfo {author}
  {\bibfnamefont {G.}~\bibnamefont {Perez}}, \ and\ \bibinfo {author}
  {\bibfnamefont {D.}~\bibnamefont {Budker}},\ }\href {\doibase
  10.1103/PhysRevLett.123.141102} {\bibfield  {journal} {\bibinfo  {journal}
  {Phys. Rev. Lett.}\ }\textbf {\bibinfo {volume} {123}},\ \bibinfo {pages}
  {141102} (\bibinfo {year} {2019})}\BibitemShut {NoStop}%
\bibitem [{\citenamefont {Arvanitaki}\ \emph {et~al.}(2018)\citenamefont
  {Arvanitaki}, \citenamefont {Graham}, \citenamefont {Hogan}, \citenamefont
  {Rajendran},\ and\ \citenamefont {Van~Tilburg}}]{Arvanitaki2018}%
  \BibitemOpen
  \bibfield  {author} {\bibinfo {author} {\bibfnamefont {A.}~\bibnamefont
  {Arvanitaki}}, \bibinfo {author} {\bibfnamefont {P.~W.}\ \bibnamefont
  {Graham}}, \bibinfo {author} {\bibfnamefont {J.~M.}\ \bibnamefont {Hogan}},
  \bibinfo {author} {\bibfnamefont {S.}~\bibnamefont {Rajendran}}, \ and\
  \bibinfo {author} {\bibfnamefont {K.}~\bibnamefont {Van~Tilburg}},\ }\href
  {\doibase 10.1103/PhysRevD.97.075020} {\bibfield  {journal} {\bibinfo
  {journal} {Phys. Rev. D}\ }\textbf {\bibinfo {volume} {97}},\ \bibinfo
  {pages} {075020} (\bibinfo {year} {2018})}\BibitemShut {NoStop}%
\bibitem [{\citenamefont {Arvanitaki}\ \emph {et~al.}(2016)\citenamefont
  {Arvanitaki}, \citenamefont {Dimopoulos},\ and\ \citenamefont {{Van
  Tilburg}}}]{Arvanitaki2016}%
  \BibitemOpen
  \bibfield  {author} {\bibinfo {author} {\bibfnamefont {A.}~\bibnamefont
  {Arvanitaki}}, \bibinfo {author} {\bibfnamefont {S.}~\bibnamefont
  {Dimopoulos}}, \ and\ \bibinfo {author} {\bibfnamefont {K.}~\bibnamefont
  {{Van Tilburg}}},\ }\href {\doibase 10.1103/PhysRevLett.116.031102}
  {\bibfield  {journal} {\bibinfo  {journal} {Phys. Rev. Lett.}\ }\textbf
  {\bibinfo {volume} {116}},\ \bibinfo {pages} {031102} (\bibinfo {year}
  {2016})},\ \Eprint {http://arxiv.org/abs/1508.01798} {arXiv:1508.01798}
  \BibitemShut {NoStop}%
\bibitem [{\citenamefont {Stadnik}\ and\ \citenamefont
  {Flambaum}(2016)}]{StadnikLasInf2015}%
  \BibitemOpen
  \bibfield  {author} {\bibinfo {author} {\bibfnamefont {Y.~V.}\ \bibnamefont
  {Stadnik}}\ and\ \bibinfo {author} {\bibfnamefont {V.~V.}\ \bibnamefont
  {Flambaum}},\ }\href {\doibase 10.1103/PhysRevA.93.063630} {\bibfield
  {journal} {\bibinfo  {journal} {Phys. Rev. A}\ }\textbf {\bibinfo {volume}
  {93}},\ \bibinfo {pages} {063630} (\bibinfo {year} {2016})}\BibitemShut
  {NoStop}%
\bibitem [{\citenamefont {{Morisaki}}\ and\ \citenamefont
  {{Suyama}}(2019)}]{morisaki:2019aa}%
  \BibitemOpen
  \bibfield  {author} {\bibinfo {author} {\bibfnamefont {S.}~\bibnamefont
  {{Morisaki}}}\ and\ \bibinfo {author} {\bibfnamefont {T.}~\bibnamefont
  {{Suyama}}},\ }\href {\doibase 10.1103/PhysRevD.100.123512} {\bibfield
  {journal} {\bibinfo  {journal} {\prd}\ }\textbf {\bibinfo {volume} {100}},\
  \bibinfo {eid} {123512} (\bibinfo {year} {2019})},\ \Eprint
  {http://arxiv.org/abs/1811.05003} {arXiv:1811.05003 [hep-ph]} \BibitemShut
  {NoStop}%
\bibitem [{\citenamefont {Grote}\ and\ \citenamefont
  {Stadnik}(2019)}]{Grote2019a}%
  \BibitemOpen
  \bibfield  {author} {\bibinfo {author} {\bibfnamefont {H.}~\bibnamefont
  {Grote}}\ and\ \bibinfo {author} {\bibfnamefont {Y.~V.}\ \bibnamefont
  {Stadnik}},\ }\href {\doibase 10.1103/PhysRevResearch.1.033187} {\bibfield
  {journal} {\bibinfo  {journal} {Phys. Rev. Research}\ }\textbf {\bibinfo
  {volume} {1}},\ \bibinfo {pages} {033187} (\bibinfo {year}
  {2019})}\BibitemShut {NoStop}%
\bibitem [{\citenamefont {Savalle}\ \emph {et~al.}(2019)\citenamefont
  {Savalle}, \citenamefont {Roberts}, \citenamefont {Frank}, \citenamefont
  {Pottie}, \citenamefont {McAllister}, \citenamefont {Dailey}, \citenamefont
  {Derevianko},\ and\ \citenamefont {Wolf}}]{Savalle_2019}%
  \BibitemOpen
  \bibfield  {author} {\bibinfo {author} {\bibfnamefont {E.}~\bibnamefont
  {Savalle}}, \bibinfo {author} {\bibfnamefont {B.~M.}\ \bibnamefont
  {Roberts}}, \bibinfo {author} {\bibfnamefont {F.}~\bibnamefont {Frank}},
  \bibinfo {author} {\bibfnamefont {P.-E.}\ \bibnamefont {Pottie}}, \bibinfo
  {author} {\bibfnamefont {B.~T.}\ \bibnamefont {McAllister}}, \bibinfo
  {author} {\bibfnamefont {C.}~\bibnamefont {Dailey}}, \bibinfo {author}
  {\bibfnamefont {A.}~\bibnamefont {Derevianko}}, \ and\ \bibinfo {author}
  {\bibfnamefont {P.}~\bibnamefont {Wolf}},\ }\href@noop {} {\enquote {\bibinfo
  {title} {Novel approaches to dark-matter detection using space-time separated
  clocks},}\ } (\bibinfo {year} {2019}),\ \Eprint
  {http://arxiv.org/abs/1902.07192} {arXiv:1902.07192 [gr-qc]} \BibitemShut
  {NoStop}%
\bibitem [{\citenamefont {Damour}\ and\ \citenamefont
  {Donoghue}(2010)}]{Damour2010}%
  \BibitemOpen
  \bibfield  {author} {\bibinfo {author} {\bibfnamefont {T.}~\bibnamefont
  {Damour}}\ and\ \bibinfo {author} {\bibfnamefont {J.~F.}\ \bibnamefont
  {Donoghue}},\ }\href {\doibase 10.1103/PhysRevD.82.084033} {\bibfield
  {journal} {\bibinfo  {journal} {Phys. Rev. D}\ }\textbf {\bibinfo {volume}
  {82}},\ \bibinfo {pages} {084033} (\bibinfo {year} {2010})}\BibitemShut
  {NoStop}%
\bibitem [{\citenamefont {{McMillan}}(2011)}]{mcmillan:2011vn}%
  \BibitemOpen
  \bibfield  {author} {\bibinfo {author} {\bibfnamefont {P.~J.}\ \bibnamefont
  {{McMillan}}},\ }\href {\doibase 10.1111/j.1365-2966.2011.18564.x} {\bibfield
   {journal} {\bibinfo  {journal} {\mnras}\ }\textbf {\bibinfo {volume}
  {414}},\ \bibinfo {pages} {2446} (\bibinfo {year} {2011})},\ \Eprint
  {http://arxiv.org/abs/1102.4340} {arXiv:1102.4340 [astro-ph.GA]} \BibitemShut
  {NoStop}%
\bibitem [{\citenamefont {Wagner}\ \emph {et~al.}(2012)\citenamefont {Wagner},
  \citenamefont {Schlamminger}, \citenamefont {Gundlach},\ and\ \citenamefont
  {Adelberger}}]{Wagner_2012}%
  \BibitemOpen
  \bibfield  {author} {\bibinfo {author} {\bibfnamefont {T.~A.}\ \bibnamefont
  {Wagner}}, \bibinfo {author} {\bibfnamefont {S.}~\bibnamefont
  {Schlamminger}}, \bibinfo {author} {\bibfnamefont {J.~H.}\ \bibnamefont
  {Gundlach}}, \ and\ \bibinfo {author} {\bibfnamefont {E.~G.}\ \bibnamefont
  {Adelberger}},\ }\href {\doibase 10.1088/0264-9381/29/18/184002} {\bibfield
  {journal} {\bibinfo  {journal} {Classical and Quantum Gravity}\ }\textbf
  {\bibinfo {volume} {29}},\ \bibinfo {pages} {184002} (\bibinfo {year}
  {2012})}\BibitemShut {NoStop}%
\bibitem [{\citenamefont {{Berg{\'e}}}\ \emph {et~al.}(2018)\citenamefont
  {{Berg{\'e}}}, \citenamefont {{Brax}}, \citenamefont {{M{\'e}tris}},
  \citenamefont {{Pernot-Borr{\`a}s}}, \citenamefont {{Touboul}},\ and\
  \citenamefont {{Uzan}}}]{berge:2018aa}%
  \BibitemOpen
  \bibfield  {author} {\bibinfo {author} {\bibfnamefont {J.}~\bibnamefont
  {{Berg{\'e}}}}, \bibinfo {author} {\bibfnamefont {P.}~\bibnamefont {{Brax}}},
  \bibinfo {author} {\bibfnamefont {G.}~\bibnamefont {{M{\'e}tris}}}, \bibinfo
  {author} {\bibfnamefont {M.}~\bibnamefont {{Pernot-Borr{\`a}s}}}, \bibinfo
  {author} {\bibfnamefont {P.}~\bibnamefont {{Touboul}}}, \ and\ \bibinfo
  {author} {\bibfnamefont {J.-P.}\ \bibnamefont {{Uzan}}},\ }\href {\doibase
  10.1103/PhysRevLett.120.141101} {\bibfield  {journal} {\bibinfo  {journal}
  {Physical Review Letters}\ }\textbf {\bibinfo {volume} {120}},\ \bibinfo
  {eid} {141101} (\bibinfo {year} {2018})},\ \Eprint
  {http://arxiv.org/abs/1712.00483} {arXiv:1712.00483 [gr-qc]} \BibitemShut
  {NoStop}%
\bibitem [{\citenamefont {Millo}\ \emph {et~al.}(2009)\citenamefont {Millo},
  \citenamefont {Magalh\~aes}, \citenamefont {Mandache}, \citenamefont
  {Le~Coq}, \citenamefont {English}, \citenamefont {Westergaard}, \citenamefont
  {Lodewyck}, \citenamefont {Bize}, \citenamefont {Lemonde},\ and\
  \citenamefont {Santarelli}}]{Millo2009}%
  \BibitemOpen
  \bibfield  {author} {\bibinfo {author} {\bibfnamefont {J.}~\bibnamefont
  {Millo}}, \bibinfo {author} {\bibfnamefont {D.~V.}\ \bibnamefont
  {Magalh\~aes}}, \bibinfo {author} {\bibfnamefont {C.}~\bibnamefont
  {Mandache}}, \bibinfo {author} {\bibfnamefont {Y.}~\bibnamefont {Le~Coq}},
  \bibinfo {author} {\bibfnamefont {E.~M.~L.}\ \bibnamefont {English}},
  \bibinfo {author} {\bibfnamefont {P.~G.}\ \bibnamefont {Westergaard}},
  \bibinfo {author} {\bibfnamefont {J.}~\bibnamefont {Lodewyck}}, \bibinfo
  {author} {\bibfnamefont {S.}~\bibnamefont {Bize}}, \bibinfo {author}
  {\bibfnamefont {P.}~\bibnamefont {Lemonde}}, \ and\ \bibinfo {author}
  {\bibfnamefont {G.}~\bibnamefont {Santarelli}},\ }\href {\doibase
  10.1103/PhysRevA.79.053829} {\bibfield  {journal} {\bibinfo  {journal} {Phys.
  Rev. A}\ }\textbf {\bibinfo {volume} {79}},\ \bibinfo {pages} {053829}
  (\bibinfo {year} {2009})}\BibitemShut {NoStop}%
\bibitem [{\citenamefont {Xie}\ \emph {et~al.}(2017)\citenamefont {Xie},
  \citenamefont {Bouchand}, \citenamefont {Nicolodi}, \citenamefont {Lours},
  \citenamefont {Alexandre},\ and\ \citenamefont {Coq}}]{Xie2017}%
  \BibitemOpen
  \bibfield  {author} {\bibinfo {author} {\bibfnamefont {X.}~\bibnamefont
  {Xie}}, \bibinfo {author} {\bibfnamefont {R.}~\bibnamefont {Bouchand}},
  \bibinfo {author} {\bibfnamefont {D.}~\bibnamefont {Nicolodi}}, \bibinfo
  {author} {\bibfnamefont {M.}~\bibnamefont {Lours}}, \bibinfo {author}
  {\bibfnamefont {C.}~\bibnamefont {Alexandre}}, \ and\ \bibinfo {author}
  {\bibfnamefont {Y.~L.}\ \bibnamefont {Coq}},\ }\href {\doibase
  10.1364/OL.42.001217} {\bibfield  {journal} {\bibinfo  {journal} {Opt.
  Lett.}\ }\textbf {\bibinfo {volume} {42}},\ \bibinfo {pages} {1217} (\bibinfo
  {year} {2017})}\BibitemShut {NoStop}%
\bibitem [{\citenamefont {Numata}\ \emph {et~al.}(2004)\citenamefont {Numata},
  \citenamefont {Kemery},\ and\ \citenamefont {Camp}}]{Numata2004}%
  \BibitemOpen
  \bibfield  {author} {\bibinfo {author} {\bibfnamefont {K.}~\bibnamefont
  {Numata}}, \bibinfo {author} {\bibfnamefont {A.}~\bibnamefont {Kemery}}, \
  and\ \bibinfo {author} {\bibfnamefont {J.}~\bibnamefont {Camp}},\ }\href
  {\doibase 10.1103/PhysRevLett.93.250602} {\bibfield  {journal} {\bibinfo
  {journal} {Phys. Rev. Lett.}\ }\textbf {\bibinfo {volume} {93}},\ \bibinfo
  {pages} {250602} (\bibinfo {year} {2004})}\BibitemShut {NoStop}%
\bibitem [{\citenamefont {Zhang}\ \emph {et~al.}(2013)\citenamefont {Zhang},
  \citenamefont {Luo}, \citenamefont {Ouyang}, \citenamefont {Deng},
  \citenamefont {Lu},\ and\ \citenamefont {Luo}}]{Zhang2013}%
  \BibitemOpen
  \bibfield  {author} {\bibinfo {author} {\bibfnamefont {J.}~\bibnamefont
  {Zhang}}, \bibinfo {author} {\bibfnamefont {Y.}~\bibnamefont {Luo}}, \bibinfo
  {author} {\bibfnamefont {B.}~\bibnamefont {Ouyang}}, \bibinfo {author}
  {\bibfnamefont {K.}~\bibnamefont {Deng}}, \bibinfo {author} {\bibfnamefont
  {Z.}~\bibnamefont {Lu}}, \ and\ \bibinfo {author} {\bibfnamefont
  {J.}~\bibnamefont {Luo}},\ }\href {\doibase 10.1140/epjd/e2013-30458-2}
  {\bibfield  {journal} {\bibinfo  {journal} {The European Physical Journal D}\
  }\textbf {\bibinfo {volume} {67}},\ \bibinfo {pages} {46} (\bibinfo {year}
  {2013})}\BibitemShut {NoStop}%
\bibitem [{sup()}]{supplemental}%
  \BibitemOpen
  \href@noop {} {}\bibinfo {note} {See Supplemental Material [url] for the
  detailed calculation of the cavity mechanical resonance and cavity optical
  resonance, for the cavity noise floor and for a detailed presentation of the
  statistical approach used in this analysis, which includes
  Ref.~\cite{Arvanitaki2016,Grote2019a,Millo2009,Xie2017,Centers2019,Foster2018,
  Derevianko2016,Banerjee2020,Aharony2019,Antypas2019,Banerjee2019,Lee2016,
  Vinet2010,Canuel2017}}\BibitemShut {NoStop}%
\bibitem [{\citenamefont {Braxmaier}\ \emph {et~al.}(2001)\citenamefont
  {Braxmaier}, \citenamefont {Pradl}, \citenamefont {M\"uller}, \citenamefont
  {Peters}, \citenamefont {Mlynek}, \citenamefont {Loriette},\ and\
  \citenamefont {Schiller}}]{Braxmaier2001}%
  \BibitemOpen
  \bibfield  {author} {\bibinfo {author} {\bibfnamefont {C.}~\bibnamefont
  {Braxmaier}}, \bibinfo {author} {\bibfnamefont {O.}~\bibnamefont {Pradl}},
  \bibinfo {author} {\bibfnamefont {H.}~\bibnamefont {M\"uller}}, \bibinfo
  {author} {\bibfnamefont {A.}~\bibnamefont {Peters}}, \bibinfo {author}
  {\bibfnamefont {J.}~\bibnamefont {Mlynek}}, \bibinfo {author} {\bibfnamefont
  {V.}~\bibnamefont {Loriette}}, \ and\ \bibinfo {author} {\bibfnamefont
  {S.}~\bibnamefont {Schiller}},\ }\href {\doibase 10.1103/PhysRevD.64.042001}
  {\bibfield  {journal} {\bibinfo  {journal} {Phys. Rev. D}\ }\textbf {\bibinfo
  {volume} {64}},\ \bibinfo {pages} {042001} (\bibinfo {year}
  {2001})}\BibitemShut {NoStop}%
\bibitem [{\citenamefont {{Scargle}}(1982)}]{Scargle1982}%
  \BibitemOpen
  \bibfield  {author} {\bibinfo {author} {\bibfnamefont {J.~D.}\ \bibnamefont
  {{Scargle}}},\ }\href {\doibase 10.1086/160554} {\bibfield  {journal}
  {\bibinfo  {journal} {\apj}\ }\textbf {\bibinfo {volume} {263}},\ \bibinfo
  {pages} {835} (\bibinfo {year} {1982})}\BibitemShut {NoStop}%
\bibitem [{\citenamefont {Freese}\ \emph {et~al.}(2013)\citenamefont {Freese},
  \citenamefont {Lisanti},\ and\ \citenamefont {Savage}}]{Freese2013}%
  \BibitemOpen
  \bibfield  {author} {\bibinfo {author} {\bibfnamefont {K.}~\bibnamefont
  {Freese}}, \bibinfo {author} {\bibfnamefont {M.}~\bibnamefont {Lisanti}}, \
  and\ \bibinfo {author} {\bibfnamefont {C.}~\bibnamefont {Savage}},\ }\href
  {\doibase 10.1103/RevModPhys.85.1561} {\bibfield  {journal} {\bibinfo
  {journal} {Rev. Mod. Phys.}\ }\textbf {\bibinfo {volume} {85}},\ \bibinfo
  {pages} {1561} (\bibinfo {year} {2013})}\BibitemShut {NoStop}%
\bibitem [{\citenamefont {Pillepich}\ \emph {et~al.}(2014)\citenamefont
  {Pillepich}, \citenamefont {Kuhlen}, \citenamefont {Guedes},\ and\
  \citenamefont {Madau}}]{Pillepich2014}%
  \BibitemOpen
  \bibfield  {author} {\bibinfo {author} {\bibfnamefont {A.}~\bibnamefont
  {Pillepich}}, \bibinfo {author} {\bibfnamefont {M.}~\bibnamefont {Kuhlen}},
  \bibinfo {author} {\bibfnamefont {J.}~\bibnamefont {Guedes}}, \ and\ \bibinfo
  {author} {\bibfnamefont {P.}~\bibnamefont {Madau}},\ }\href {\doibase
  10.1088/0004-637x/784/2/161} {\bibfield  {journal} {\bibinfo  {journal} {The
  Astrophysical Journal}\ }\textbf {\bibinfo {volume} {784}},\ \bibinfo {pages}
  {161} (\bibinfo {year} {2014})}\BibitemShut {NoStop}%
\bibitem [{\citenamefont {Derevianko}(2018{\natexlab{a}})}]{Derevianko2016}%
  \BibitemOpen
  \bibfield  {author} {\bibinfo {author} {\bibfnamefont {A.}~\bibnamefont
  {Derevianko}},\ }\href {\doibase 10.1103/PhysRevA.97.042506} {\bibfield
  {journal} {\bibinfo  {journal} {Phys. Rev. A}\ }\textbf {\bibinfo {volume}
  {97}},\ \bibinfo {pages} {042506} (\bibinfo {year}
  {2018}{\natexlab{a}})}\BibitemShut {NoStop}%
\bibitem [{\citenamefont {Foster}\ \emph {et~al.}(2018)\citenamefont {Foster},
  \citenamefont {Rodd},\ and\ \citenamefont {Safdi}}]{Foster2018}%
  \BibitemOpen
  \bibfield  {author} {\bibinfo {author} {\bibfnamefont {J.~W.}\ \bibnamefont
  {Foster}}, \bibinfo {author} {\bibfnamefont {N.~L.}\ \bibnamefont {Rodd}}, \
  and\ \bibinfo {author} {\bibfnamefont {B.~R.}\ \bibnamefont {Safdi}},\ }\href
  {\doibase 10.1103/PhysRevD.97.123006} {\bibfield  {journal} {\bibinfo
  {journal} {Phys. Rev. D}\ }\textbf {\bibinfo {volume} {97}},\ \bibinfo
  {pages} {123006} (\bibinfo {year} {2018})}\BibitemShut {NoStop}%
\bibitem [{\citenamefont {{Evans}}\ \emph {et~al.}(2019)\citenamefont
  {{Evans}}, \citenamefont {{O'Hare}},\ and\ \citenamefont
  {{McCabe}}}]{evans:2019aa}%
  \BibitemOpen
  \bibfield  {author} {\bibinfo {author} {\bibfnamefont {N.~W.}\ \bibnamefont
  {{Evans}}}, \bibinfo {author} {\bibfnamefont {C.~A.~J.}\ \bibnamefont
  {{O'Hare}}}, \ and\ \bibinfo {author} {\bibfnamefont {C.}~\bibnamefont
  {{McCabe}}},\ }\href {\doibase 10.1103/PhysRevD.99.023012} {\bibfield
  {journal} {\bibinfo  {journal} {\prd}\ }\textbf {\bibinfo {volume} {99}},\
  \bibinfo {eid} {023012} (\bibinfo {year} {2019})}\BibitemShut {NoStop}%
\bibitem [{\citenamefont
  {Derevianko}(2018{\natexlab{b}})}]{DereviankoVULF2016}%
  \BibitemOpen
  \bibfield  {author} {\bibinfo {author} {\bibfnamefont {A.}~\bibnamefont
  {Derevianko}},\ }\href {\doibase 10.1103/PhysRevA.97.042506} {\bibfield
  {journal} {\bibinfo  {journal} {Phys. Rev. A}\ }\textbf {\bibinfo {volume}
  {97}},\ \bibinfo {pages} {042506} (\bibinfo {year}
  {2018}{\natexlab{b}})}\BibitemShut {NoStop}%
\bibitem [{\citenamefont {Centers}\ \emph {et~al.}(2019)\citenamefont
  {Centers}, \citenamefont {Blanchard}, \citenamefont {Conrad}, \citenamefont
  {Figueroa}, \citenamefont {Garcon}, \citenamefont {Gramolin}, \citenamefont
  {Kimball}, \citenamefont {Lawson}, \citenamefont {Pelssers}, \citenamefont
  {Smiga}, \citenamefont {Sushkov}, \citenamefont {Wickenbrock}, \citenamefont
  {Budker},\ and\ \citenamefont {Derevianko}}]{Centers2019}%
  \BibitemOpen
  \bibfield  {author} {\bibinfo {author} {\bibfnamefont {G.~P.}\ \bibnamefont
  {Centers}}, \bibinfo {author} {\bibfnamefont {J.~W.}\ \bibnamefont
  {Blanchard}}, \bibinfo {author} {\bibfnamefont {J.}~\bibnamefont {Conrad}},
  \bibinfo {author} {\bibfnamefont {N.~L.}\ \bibnamefont {Figueroa}}, \bibinfo
  {author} {\bibfnamefont {A.}~\bibnamefont {Garcon}}, \bibinfo {author}
  {\bibfnamefont {A.~V.}\ \bibnamefont {Gramolin}}, \bibinfo {author}
  {\bibfnamefont {D.~F.~J.}\ \bibnamefont {Kimball}}, \bibinfo {author}
  {\bibfnamefont {M.}~\bibnamefont {Lawson}}, \bibinfo {author} {\bibfnamefont
  {B.}~\bibnamefont {Pelssers}}, \bibinfo {author} {\bibfnamefont {J.~A.}\
  \bibnamefont {Smiga}}, \bibinfo {author} {\bibfnamefont {A.~O.}\ \bibnamefont
  {Sushkov}}, \bibinfo {author} {\bibfnamefont {A.}~\bibnamefont
  {Wickenbrock}}, \bibinfo {author} {\bibfnamefont {D.}~\bibnamefont {Budker}},
  \ and\ \bibinfo {author} {\bibfnamefont {A.}~\bibnamefont {Derevianko}},\
  }\href@noop {} {\enquote {\bibinfo {title} {Stochastic fluctuations of
  bosonic dark matter},}\ } (\bibinfo {year} {2019}),\ \Eprint
  {http://arxiv.org/abs/1905.13650} {arXiv:1905.13650 [astro-ph.CO]}
  \BibitemShut {NoStop}%
\bibitem [{\citenamefont {Schlamminger}\ \emph {et~al.}(2008)\citenamefont
  {Schlamminger}, \citenamefont {Choi}, \citenamefont {Wagner}, \citenamefont
  {Gundlach},\ and\ \citenamefont {Adelberger}}]{Schlamminger2008}%
  \BibitemOpen
  \bibfield  {author} {\bibinfo {author} {\bibfnamefont {S.}~\bibnamefont
  {Schlamminger}}, \bibinfo {author} {\bibfnamefont {K.-Y.}\ \bibnamefont
  {Choi}}, \bibinfo {author} {\bibfnamefont {T.~A.}\ \bibnamefont {Wagner}},
  \bibinfo {author} {\bibfnamefont {J.}~\bibnamefont {Gundlach}}, \ and\
  \bibinfo {author} {\bibfnamefont {E.~G.}\ \bibnamefont {Adelberger}},\
  }\href@noop {} {\bibfield  {journal} {\bibinfo  {journal} {Phys. Rev. Lett}\
  }\textbf {\bibinfo {volume} {100}},\ \bibinfo {pages} {041101} (\bibinfo
  {year} {2008})}\BibitemShut {NoStop}%
\bibitem [{\citenamefont {Banerjee}\ \emph {et~al.}(2020)\citenamefont
  {Banerjee}, \citenamefont {Budker}, \citenamefont {Eby}, \citenamefont
  {Kim},\ and\ \citenamefont {Perez}}]{Banerjee2020}%
  \BibitemOpen
  \bibfield  {author} {\bibinfo {author} {\bibfnamefont {A.}~\bibnamefont
  {Banerjee}}, \bibinfo {author} {\bibfnamefont {D.}~\bibnamefont {Budker}},
  \bibinfo {author} {\bibfnamefont {J.}~\bibnamefont {Eby}}, \bibinfo {author}
  {\bibfnamefont {H.}~\bibnamefont {Kim}}, \ and\ \bibinfo {author}
  {\bibfnamefont {G.}~\bibnamefont {Perez}},\ }\href {\doibase
  10.1038/s42005-019-0260-3} {\bibfield  {journal} {\bibinfo  {journal}
  {Communications Physics}\ }\textbf {\bibinfo {volume} {3}},\ \bibinfo {pages}
  {1} (\bibinfo {year} {2020})}\BibitemShut {NoStop}%
\bibitem [{\citenamefont {Kolb}\ and\ \citenamefont
  {Tkachev}(1993)}]{Kolb1993}%
  \BibitemOpen
  \bibfield  {author} {\bibinfo {author} {\bibfnamefont {E.~W.}\ \bibnamefont
  {Kolb}}\ and\ \bibinfo {author} {\bibfnamefont {I.~I.}\ \bibnamefont
  {Tkachev}},\ }\href {\doibase 10.1103/physrevlett.71.3051} {\bibfield
  {journal} {\bibinfo  {journal} {Physical Review Letters}\ }\textbf {\bibinfo
  {volume} {71}},\ \bibinfo {pages} {3051} (\bibinfo {year}
  {1993})}\BibitemShut {NoStop}%
\bibitem [{\citenamefont {Levkov}\ \emph {et~al.}(2018)\citenamefont {Levkov},
  \citenamefont {Panin},\ and\ \citenamefont {Tkachev}}]{Levkov2018}%
  \BibitemOpen
  \bibfield  {author} {\bibinfo {author} {\bibfnamefont {D.~G.}\ \bibnamefont
  {Levkov}}, \bibinfo {author} {\bibfnamefont {A.~G.}\ \bibnamefont {Panin}}, \
  and\ \bibinfo {author} {\bibfnamefont {I.~I.}\ \bibnamefont {Tkachev}},\
  }\href {\doibase 10.1103/PhysRevLett.121.151301} {\bibfield  {journal}
  {\bibinfo  {journal} {Phys. Rev. Lett.}\ }\textbf {\bibinfo {volume} {121}},\
  \bibinfo {pages} {151301} (\bibinfo {year} {2018})}\BibitemShut {NoStop}%
\bibitem [{\citenamefont {Braaten}\ and\ \citenamefont
  {Zhang}(2019)}]{Braaten2019}%
  \BibitemOpen
  \bibfield  {author} {\bibinfo {author} {\bibfnamefont {E.}~\bibnamefont
  {Braaten}}\ and\ \bibinfo {author} {\bibfnamefont {H.}~\bibnamefont
  {Zhang}},\ }\href {\doibase 10.1103/RevModPhys.91.041002} {\bibfield
  {journal} {\bibinfo  {journal} {Rev. Mod. Phys.}\ }\textbf {\bibinfo {volume}
  {91}},\ \bibinfo {pages} {041002} (\bibinfo {year} {2019})}\BibitemShut
  {NoStop}%
\bibitem [{\citenamefont {Vaquero}\ \emph {et~al.}(2019)\citenamefont
  {Vaquero}, \citenamefont {Redondo},\ and\ \citenamefont
  {Stadler}}]{Vaquero2019}%
  \BibitemOpen
  \bibfield  {author} {\bibinfo {author} {\bibfnamefont {A.}~\bibnamefont
  {Vaquero}}, \bibinfo {author} {\bibfnamefont {J.}~\bibnamefont {Redondo}}, \
  and\ \bibinfo {author} {\bibfnamefont {J.}~\bibnamefont {Stadler}},\ }\href
  {\doibase 10.1088/1475-7516/2019/04/012} {\bibfield  {journal} {\bibinfo
  {journal} {Journal of Cosmology and Astroparticle Physics}\ }\textbf
  {\bibinfo {volume} {2019}},\ \bibinfo {pages} {012} (\bibinfo {year}
  {2019})}\BibitemShut {NoStop}%
\bibitem [{\citenamefont {Bar-Or}\ \emph {et~al.}(2019)\citenamefont {Bar-Or},
  \citenamefont {Fouvry},\ and\ \citenamefont {Tremaine}}]{BarOr2019}%
  \BibitemOpen
  \bibfield  {author} {\bibinfo {author} {\bibfnamefont {B.}~\bibnamefont
  {Bar-Or}}, \bibinfo {author} {\bibfnamefont {J.-B.}\ \bibnamefont {Fouvry}},
  \ and\ \bibinfo {author} {\bibfnamefont {S.}~\bibnamefont {Tremaine}},\
  }\href {\doibase 10.3847/1538-4357/aaf28c} {\bibfield  {journal} {\bibinfo
  {journal} {The Astrophysical Journal}\ }\textbf {\bibinfo {volume} {871}},\
  \bibinfo {pages} {28} (\bibinfo {year} {2019})}\BibitemShut {NoStop}%
\bibitem [{\citenamefont {Aharony}\ \emph {et~al.}(2019)\citenamefont
  {Aharony}, \citenamefont {Akerman}, \citenamefont {Ozeri}, \citenamefont
  {Perez}, \citenamefont {Savoray},\ and\ \citenamefont
  {Shaniv}}]{Aharony2019}%
  \BibitemOpen
  \bibfield  {author} {\bibinfo {author} {\bibfnamefont {S.}~\bibnamefont
  {Aharony}}, \bibinfo {author} {\bibfnamefont {N.}~\bibnamefont {Akerman}},
  \bibinfo {author} {\bibfnamefont {R.}~\bibnamefont {Ozeri}}, \bibinfo
  {author} {\bibfnamefont {G.}~\bibnamefont {Perez}}, \bibinfo {author}
  {\bibfnamefont {I.}~\bibnamefont {Savoray}}, \ and\ \bibinfo {author}
  {\bibfnamefont {R.}~\bibnamefont {Shaniv}},\ }\href
  {http://arxiv.org/abs/1902.02788} {\bibfield  {journal} {\bibinfo  {journal}
  {arXiv:1902.02788 [hep-ph, physics:physics]}\ } (\bibinfo {year} {2019})},\
  \bibinfo {note} {arXiv: 1902.02788}\BibitemShut {NoStop}%
\bibitem [{\citenamefont {Lee}(2016)}]{Lee2016}%
  \BibitemOpen
  \bibfield  {author} {\bibinfo {author} {\bibfnamefont {Y.-J.}\ \bibnamefont
  {Lee}},\ }\href
  {https://ocw.mit.edu/courses/physics/8-03sc-physics-iii-vibrations-and-waves-fall-2016/part-i-mechanical-vibrations-and-waves/}
  {\emph {\bibinfo {title} {Physics III: Vibrations and Waves. Part 1:
  Mechanical Vibrations and Waves}}},\ edited by\ \bibinfo {editor}
  {\bibnamefont {MIT}},\ \bibinfo {number} {8.03SC}\ (\bibinfo  {publisher}
  {Massachusetts Institute of Technology},\ \bibinfo {year} {2016})\BibitemShut
  {NoStop}%
\bibitem [{\citenamefont {Virgo-collaboration}(2010)}]{Vinet2010}%
  \BibitemOpen
  \bibfield  {author} {\bibinfo {author} {\bibnamefont {Virgo-collaboration}},\
  }\href
  {https://www.ego-gw.it/public/events/vesf/2010/Presentations/Interferometer_Materials-Vinet.pdf}
  {\emph {\bibinfo {title} {The VIRGO Physics Book, Vol. II, OPTICS and related
  TOPICS}}}\ (\bibinfo  {publisher} {The Virgo collaboration},\ \bibinfo {year}
  {2010})\BibitemShut {NoStop}%
\bibitem [{\citenamefont {{Canuel}}\ \emph {et~al.}(2018)\citenamefont
  {{Canuel}}, \citenamefont {{Bertoldi}}, \citenamefont {{Amand}},
  \citenamefont {{Borgo di Pozzo}}, \citenamefont {{Fang}}, \citenamefont
  {{Geiger}}, \citenamefont {{Gillot}}, \citenamefont {{Henry}}, \citenamefont
  {{Hinderer}}, \citenamefont {{Holleville}}, \citenamefont {{Lef{\`e}vre}},
  \citenamefont {{Merzougui}}, \citenamefont {{Mielec}}, \citenamefont
  {{Monfret}}, \citenamefont {{Pelisson}}, \citenamefont {{Prevedelli}},
  \citenamefont {{Reynaud}}, \citenamefont {{Riou}}, \citenamefont
  {{Rogister}}, \citenamefont {{Rosat}}, \citenamefont {{Cormier}},
  \citenamefont {{Landragin}}, \citenamefont {{Chaibi}}, \citenamefont
  {{Gaffet}},\ and\ \citenamefont {{Bouyer}}}]{Canuel2017}%
  \BibitemOpen
  \bibfield  {author} {\bibinfo {author} {\bibfnamefont {B.}~\bibnamefont
  {{Canuel}}}, \bibinfo {author} {\bibfnamefont {A.}~\bibnamefont
  {{Bertoldi}}}, \bibinfo {author} {\bibfnamefont {L.}~\bibnamefont {{Amand}}},
  \bibinfo {author} {\bibfnamefont {E.}~\bibnamefont {{Borgo di Pozzo}}},
  \bibinfo {author} {\bibfnamefont {B.}~\bibnamefont {{Fang}}}, \bibinfo
  {author} {\bibfnamefont {R.}~\bibnamefont {{Geiger}}}, \bibinfo {author}
  {\bibfnamefont {J.}~\bibnamefont {{Gillot}}}, \bibinfo {author}
  {\bibfnamefont {S.}~\bibnamefont {{Henry}}}, \bibinfo {author} {\bibfnamefont
  {J.}~\bibnamefont {{Hinderer}}}, \bibinfo {author} {\bibfnamefont
  {D.}~\bibnamefont {{Holleville}}}, \bibinfo {author} {\bibfnamefont
  {G.}~\bibnamefont {{Lef{\`e}vre}}}, \bibinfo {author} {\bibfnamefont
  {M.}~\bibnamefont {{Merzougui}}}, \bibinfo {author} {\bibfnamefont
  {N.}~\bibnamefont {{Mielec}}}, \bibinfo {author} {\bibfnamefont
  {T.}~\bibnamefont {{Monfret}}}, \bibinfo {author} {\bibfnamefont
  {S.}~\bibnamefont {{Pelisson}}}, \bibinfo {author} {\bibfnamefont
  {M.}~\bibnamefont {{Prevedelli}}}, \bibinfo {author} {\bibfnamefont
  {S.}~\bibnamefont {{Reynaud}}}, \bibinfo {author} {\bibfnamefont
  {I.}~\bibnamefont {{Riou}}}, \bibinfo {author} {\bibfnamefont
  {Y.}~\bibnamefont {{Rogister}}}, \bibinfo {author} {\bibfnamefont
  {S.}~\bibnamefont {{Rosat}}}, \bibinfo {author} {\bibfnamefont
  {E.}~\bibnamefont {{Cormier}}}, \bibinfo {author} {\bibfnamefont
  {A.}~\bibnamefont {{Landragin}}}, \bibinfo {author} {\bibfnamefont
  {W.}~\bibnamefont {{Chaibi}}}, \bibinfo {author} {\bibfnamefont
  {S.}~\bibnamefont {{Gaffet}}}, \ and\ \bibinfo {author} {\bibfnamefont
  {P.}~\bibnamefont {{Bouyer}}},\ }\href {\doibase 10.1038/s41598-018-32165-z}
  {\bibfield  {journal} {\bibinfo  {journal} {Scientific Reports}\ }\textbf
  {\bibinfo {volume} {8}},\ \bibinfo {pages} {14064} (\bibinfo {year}
  {2018})},\ \Eprint {http://arxiv.org/abs/1703.02490} {arXiv:1703.02490
  [physics.atom-ph]} \BibitemShut {NoStop}%
\bibitem [{\citenamefont {Banerjee}\ \emph {et~al.}()\citenamefont {Banerjee},
  \citenamefont {Budker}, \citenamefont {Eby}, \citenamefont {Flambaum},
  \citenamefont {Kim}, \citenamefont {Matsedonskyi},\ and\ \citenamefont
  {Perez}}]{Banerjee2019}%
  \BibitemOpen
  \bibfield  {author} {\bibinfo {author} {\bibfnamefont {A.}~\bibnamefont
  {Banerjee}}, \bibinfo {author} {\bibfnamefont {D.}~\bibnamefont {Budker}},
  \bibinfo {author} {\bibfnamefont {J.}~\bibnamefont {Eby}}, \bibinfo {author}
  {\bibfnamefont {V.~V.}\ \bibnamefont {Flambaum}}, \bibinfo {author}
  {\bibfnamefont {H.}~\bibnamefont {Kim}}, \bibinfo {author} {\bibfnamefont
  {O.}~\bibnamefont {Matsedonskyi}}, \ and\ \bibinfo {author} {\bibfnamefont
  {G.}~\bibnamefont {Perez}},\ }\href@noop {} {\ }\Eprint
  {http://arxiv.org/abs/1912.04295} {arXiv:1912.04295 [hep-ph]} \BibitemShut
  {NoStop}%
\end{thebibliography}%
\clearpage

\section{Supplemental material A : Cavity mechanical resonance}\label{App:A}

In this appendix we model the resonant cavity in the presence of a temporal oscillation of the fundamental constants. As shown in \cite{Arvanitaki2016,Grote2019a} using a simple ``mass-spring'' model, the effect of the Bohr radius change is a ``driving'' force of the harmonic oscillator whose equation of motion is then
\begin{equation} \label{eq:harmonic_osc_D2}
\ddot{D}(t) + \frac{\omega_{r}}{Q_0}\dot{D}(t) + \omega_{r}^2 D(t) =-\epsilon_L L_0\, \omega_m^2 \cos(\omega_m t) \, ,
\end{equation}
where $\omega_r$ is the resonant frequency and $Q_0$ its quality factor. We define the displacement $D(t) \equiv L(t)-L_{eq}(t)$ where $L(t)$ is the cavity length and $L_{eq}(t) \equiv  L_{0}(1-\epsilon_L \cos(\omega_mt))$ the equilibrium length. It is deviations with respect to $L_{eq}(t)$ that give rise to restoring and damping forces.

The simple mass-spring model can be generalised to an elastic solid cavity using the standard methods described in e.g. \cite{Lee2016}. The harmonic oscillator (\ref{eq:harmonic_osc_D2}) becomes a wave equation for the function $D(t,x)$ representing the displacement with respect to the (time varying) equilibrium position of any segment at position $x$ (we choose $x=0$ at the cavity centre):
\begin{equation}\label{eq:wave_eq}
\ddot{D}(t,x) -\frac{\partial^2}{\partial x^2}\left(\kappa\, D(t,x) +\gamma\,\dot{D}(t,x)\right) = -x\epsilon_L\omega_m^2\mathrm{cos}(\omega_mt)\,,
\end{equation}
where $\kappa, \gamma$ are material dependent constants.

Boundary conditions (free ends) impose the spatial modes $u^{(n)}(x)$ of form 
\begin{equation}\label{eq:spatial_modes}
u^{(n)}(x) = \sqrt{\frac{2}{L_0}}\mathrm{cos}\left(\frac{n\pi}{L_0}\left(x+\frac{L_0}{2}\right)\right) \,,
\end{equation}
where $n$ is an integer. The steady state solution is then given by a superposition of those modes and can be written as $D(t,x) = \sum_{n=1}^\infty D^{(n)}(t)\,u^{(n)}(x)$, where the $D^{(n)}(t)$ must oscillate at $\omega_m$ and satisfy
\begin{eqnarray}\label{eq:monty_modes}
\ddot{D}^{(n)}(t) +\frac{\omega_n}{Q_n}\dot{D}^{(n)}(t) +\omega_n^2 D^{(n)}(t) &=& -\epsilon_L\omega_m^2 \mathrm{cos}(\omega_mt)\int_{-L_0/2}^{L_0/2} x\,u^{(n)}(x)\, dx\, \nonumber \\
&=& b^{(n)}\epsilon_L L_0\,\omega_m^2 \mathrm{cos}(\omega_mt) \,,
\end{eqnarray} 
where $\omega_n$ and $Q_n$ are material dependent constants: $\omega_n = n v_s\pi/L_0 \approx n\,173$~krad/s ($v_s=\sqrt{\kappa}$ is the phase velocity of longitudinal elastic waves in ULE) and $Q_n = \frac{\kappa}{\gamma\omega_n}=\frac{Q_1}{n}$ with $Q_1 \approx 6.1\times 10^4$. The factor $b^{(n)} =\frac{(2)^{3/2}\sqrt{L_0}}{n^2\pi^2}$ for odd $n$, and zero for even $n$. So only modes with odd $n$ are excited as one would expect from the symmetry of the driving force. Equations (\ref{eq:monty_modes}) have analytical solutions giving the final result
\begin{equation}\label{eq:monty_L_final}
L(t) = L_0\left(1-\epsilon_L\left((1+\alpha)\,\mathrm{cos}(\omega_mt)+\beta \,\mathrm{sin}(\omega_mt)\right)\right) \,,
\end{equation}
with
\begin{equation}\label{eq:monty_finalab}
\alpha = \sum_{i=1}^\infty \frac{8}{n^2\pi^2}\frac{Q_n^2 \omega_{m}^2 \left(\omega_n^2-\omega_{m}^2\right)}{Q_n^2 \left(\omega_n^2-\omega_{m}^2\right)^2+\omega_n^2 \omega_{m}^2} \hspace{1 cm} \beta = \sum_{i=1}^\infty \frac{8}{n^2\pi^2}\frac{Q_n \omega_n \omega_{m}^3}{Q_n^2 \left(\omega_n^2-\omega_{m}^2\right)^2+\omega_n^2 \omega_{m}^2} \, ,
\end{equation}
where $n=2i-1$. The sums in (\ref{eq:monty_finalab}) can be evaluated with a limited number of terms, as for DAMNED we are interested in the frequency region up to about $n=7$ and the contribution of higher resonances quickly decreases.

Below resonance ($\omega_m \ll \omega_1$) both $\beta,\alpha \simeq 0$ and the cavity length follows $L_{eq}(t)$ and the Bohr radius change. Above resonance ($\omega_m \gg \omega_1$) the coefficients converge to $\beta=0$ and $\alpha=-1$, meaning the cavity can no longer follow the oscillations of the equilibrium length.

\section{Supplemental material B : Cavity optical resonance}\label{App:B}

The description of the resonant light field inside a Fabry-Perot cavity of oscillating length $L(t) = L_0 \cos(\omega_m t)$ has been treated extensively in the context of gravitational wave detectors like LIGO, Virgo, and more recently MIGA and described in detail in e.g. \cite{Vinet2010,Canuel2017}. Those analyses apply directly to our cavity and we only recall the main results, for details the reader is referred to the original papers.

We follow in particular the analysis in annex A of \cite{Canuel2017}, starting from equ. (35) of \cite{Canuel2017}, which gives the phase variation of the resonant light field exiting a cavity whose length is varying as $L(t) =  \zeta_c L_0 \cos(\omega_m t)$ (with $\zeta_c \ll 1$),
\begin{equation} \label{eq:phi(t)}
\begin{aligned}
\phi(t) \simeq \frac{2\zeta_c L_0 \omega_0 r^2}{c\, (r^4-2r^2\cos(2\nu)+1)} &\left((r^2-1)\cos(\nu)\cos(\omega_mt) \right.\\
 & \left. -(r^2+1)\sin(\nu)\sin(\omega_mt)\right)\, ,
\end{aligned}
\end{equation}
where $r$ is the reflection coefficient of the cavity mirrors and $\nu \equiv \omega_m L_0/c$. For our cavity with finesse $\mathcal{F} \approx 800000$ \cite{Millo2009} we have $1-r^2 \approx 4\times 10^{-6}$ $\left(\frac{r^2}{1-r^2} \simeq \mathcal{F}/\pi\right)$ and $\nu \approx [2,40] \times 10^{-5}$ for our frequency range of $[10,200]$~kHz, so we will neglect the first term in (\ref{eq:phi(t)}).

The fractional frequency variation ($\delta\omega(t)/\omega_0 = \dot{\phi}(t)/\omega_0$) is given by 
\begin{equation} \label{eq:cavity_cos}
\begin{aligned}
\frac{\delta\omega(t)}{\omega_0} = \frac{-2 \zeta_c \, \nu \, r^2(1+r^2)\sin(\nu)}{r^4-2r^2\cos(2\nu)+1} &\cos(\omega_mt)\, .
\end{aligned}
\end{equation}

The result for $L(t) =  \zeta_s L_0 \sin(\omega_m t)$ is simply obtained from (\ref{eq:cavity_cos}) by shifting $\omega_mt \rightarrow \omega_mt-\pi/2$ i.e. replacing $\cos(\omega_mt) \rightarrow \sin(\omega_mt)$ and $\zeta_c \rightarrow \zeta_s$.

We identify $\zeta_c = -\epsilon_L(1+\alpha)$ and $\zeta_s = -\epsilon_L\beta$, and comparing to equation (1) of the main part of the paper we finally obtain
\begin{equation} \label{eq:Es-Ec}
\mathcal{E}_c=\mathcal{E}_s = \frac{2 \, \nu \, r^2(1+r^2)\sin(\nu)}{r^4-2r^2\cos(2\nu)+1} \, .
\end{equation}

Evaluating (\ref{eq:Es-Ec}) for our cavity and frequency range we have $\mathcal{E}_c,\mathcal{E}_s \in [0.991,0.99998]$ i.e. $\approx 1$.

\section{Supplemental material C : Cavity noise floor}
Although the cavity was characterized in detail in \cite{Xie2017}, the noise floor changed due to an air conditioning failure in the lab. As a consequence it became the limiting noise source of our experiment. For comparison, the noise of the ``reference'' interferometer (insensitive to cavity noise) was shot noise dominated with a flat PSD at $\approx 2\times 10^{-9}$~rad$^2$/Hz, almost an order of magnitude below the ``signal'' arm noise. Modelling the cavity noise floor is then necessary to estimate the constraints on the parameters since the signal-to-noise ratio has to be constructed for Bayesian analysis. 

\begin{figure}[h!]
\includegraphics[width=0.9\textwidth]{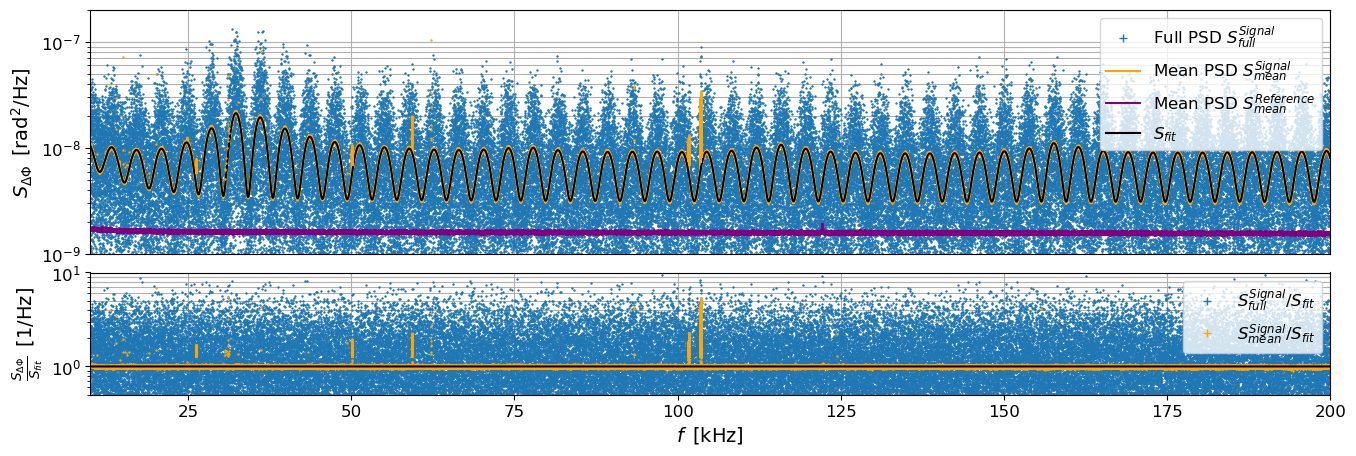}
\vspace{-0.5cm}
\caption{PSD of phase fluctuations $S_{\Delta \Phi}(f)$ over the 12 days acquisition computed using small segment of $268s$ in blue. The orange line represents the average over all the $268s$-segment PSD which is used to fit the cavity noise-floor model shown as a black line. The purple line represents the "Reference" data which are characterized by the photodiode shot noise. The bottom plot shows full and mean PSDs normalized by the model.}
\label{fig_BLUE_52_LF}
\end{figure}

The unequal-arm interferometer allows us to compare the signal from the cavity to itself after a delay corresponding to the propagation time $T$ of a photon in the fibre spools. The cavity noise $\Phi_{cav}$ creates an interferometer phase noise $\Delta \Phi_{cav}(t) = \Phi_{cav}(t) - \Phi_{cav}(t-T)$. The cavity noise PSD $S_{[\Phi_{cav}]}$ can be linked to the interferometric cavity-noise floor $S_{[\Delta \Phi_{cav}]}$ :
\begin{equation}\label{eq:PSD_appendix}
S_{[\Delta \Phi_{cav}]}(f) = 4\sin^2 \left(\pi f T\right) S_{[\Phi_{cav}]}(f) 
\end{equation}
The $4\sin^2(\pi f T)$ transfer function is responsible for the extinctions of the PSD when $f = n/T$ as seen in figure \ref{fig_BLUE_52_LF}.
In order to obtain the full PSD, we had to split the dataset in $\sim268$~s-long subsets due to RAM-limitation. In doing so, we calculate the PSD for each subset of data and ensure that the cavity noise level has not changed over the duration of the acquisition. Thanks to the stationarity of the noise, we can average these PSDs to model the characteristic noise floor of the cavity. This averaged PSD is shown in orange in figure \ref{fig_BLUE_52_LF}. The peaks are the systematic effects discussed in the main section. The model fitted to the averaged PSD is shown in black and incorporates only broad trends so as not to adjust for potential DM traces or systematic effects. Some of these systematic effects were detected using the "Reference" data, which are shown in purple.

\section{Supplemental material D : Systematic effects}
The peaks above the detection threshold in figure 2 of the main article can be broken down into two groups. The first one includes the common peaks between the ``signal" and ``reference" branches and are therefore considered systematic effects as the ``reference" branch is not long enough to be sensitive to DM. This is the case for all peaks located at a multiple of $\sim 7$~kHz and whose width is Fourier limited at 3~mHz (in red on the figure). After investigation, the origin of this signal lies in the phase-meter and is probably a digitalisation effect. The second group consists of peaks present only in the ``signal" data, which are therefore potential DM traces. This is the case of the peak around $\sim 103$~kHz. 

Four other peaks exist, in the signal data only, below the detection threshold, but clearly visible at $\sim$26, $\sim$50 ,$\sim$59 and $\sim$101~kHz when investigated in more details. The first three all have a Gaussian profile with a typical width of $\sim$~Hz. This is much too large for a putative DM signal whose width should be $\sim 10^{-6}\omega_m/2\pi \leq 0.02$~Hz. Furthermore, the mean frequency of the peaks drifts coherently over the 12 days which allows us to assume that their physical origin is the same. The other cavities available in the laboratory \cite{Xie2017} do not show peaks at the same locations even when the acquisition is performed in parallel with our main source for the experiment. 
The absence of common peaks between the cavities, expected if the signal is induced by DM, allows us to deduce that these three peaks are due to systematics. Although the exact technical origin of the peaks are unknown, we are convinced of their common physical origin and they are likely coming from the laser/cavity. The profile of the two other peaks (at $\sim101$ and $\sim103$~kHz) is also Gaussian with a width of $\sim 10$~Hz and the evolution of the peak position is correlated with the room temperature. In this frequency range, the temperature dependence and the presence of ``double" peaks strongly suggest that this signal is due to resonances in the piezoelectric block used to control the laser frequency. 

\section{Supplemental material E : stochastic modeling of the signal and data analysis}
We describe the data analysis used, based on \cite{Centers2019,Foster2018,Derevianko2016}.	
	
\subsection{Dark Matter velocity and frequency distribution}
\begin{figure}[h!]
\includegraphics[width=0.5\textwidth]{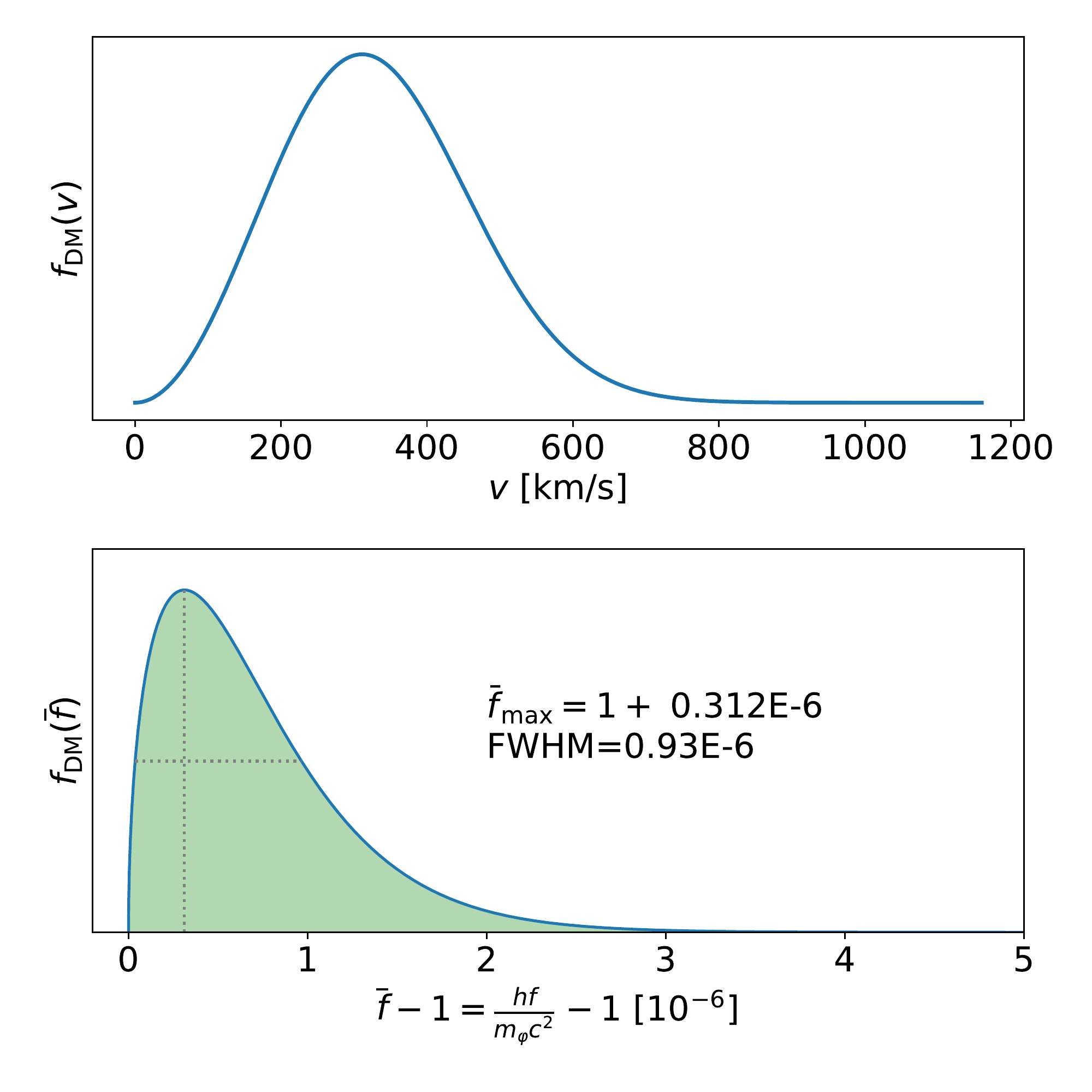}
\vspace{-0.5cm}
\caption{Top: DM velocity distribution from Eq.~(\ref{eq:fDMv}). Bottom: DM frequency distribution from Eq.~(\ref{eq:ff}). The green filled area has a width of 3 FWHM and its range is given by Eq.~(\ref{eq:rangef}) with $a=3$. This is the frequency domain over which the scalar field is modeled in Eq.~(\ref{eq:varphi}).}
\label{fig:fDM}
\end{figure}

We assume that the DM velocity distribution has a Gaussian profile characterized by a central velocity $\bm v_\mathrm{obs}$, the galactic velocity of the Solar System, and by a virial velocity $\sigma_v$ \footnote{Most of this section is presented within the standard galactic DM model, but its application to the Earth relaxion halo model is straightforward as discussed in the last subsection.}
\begin{equation}
	f_\mathrm{DM}(\bm v) = \frac{1}{(2\pi \sigma_v^2)^{3/2}}e^{-\frac{(\bm v-v_\mathrm{obs})^2}{2\sigma_v^2}} \, .
\end{equation}
This leads to a distribution in term of velocity amplitude $v$ given by
\begin{equation}\label{eq:fDMv}
	f_\mathrm{DM}(v)=\sqrt{\frac{2}{\pi}} \frac{v}{ v_\mathrm{obs} \sigma_v}e^{-\frac{v^2+ v_\mathrm{obs}^2}{2\sigma_v^2}}\sinh\left(\frac{v  v_\mathrm{obs}}{\sigma_v^2}\right) \, .
\end{equation}
Typical values for the two velocity parameters are: $v_\mathrm{obs}\sim 230$ km/s and $\sigma_v\sim 150$ km/s \cite{Foster2018}. The top panel of Fig.~\ref{fig:fDM} shows this velocity distribution.

Using the fact that the Compton frequency for the scalar field is related to the DM velocity through
\begin{equation}\label{eq:f}
	f = \frac{m_\varphi c^2}{h}\left(1+\frac{v^2}{2c^2}\right) \, ,
\end{equation}
the DM velocity distribution can be transformed into a frequency distribution \cite{Derevianko2016,Foster2018}
		\begin{align}\label{eq:ff}
f_\mathrm{DM}(f)=\frac{h}{m_{\varphi}  c^2}\sqrt{\frac{2}{\pi}}\frac{c^2}{v_\mathrm{obs}\sigma_v} e^{-\frac{2c^2(\bar f -1)+ v_\mathrm{obs}^2}{2\sigma_v^2}}\sinh\left(\frac{c  v_\mathrm{obs}}{\sigma_v^2}\sqrt{2(\bar f-1)}\right)\, ,
		\end{align}
where $\bar f$ is a dimensionless frequency defined as
\begin{equation}
	\bar f=\frac{fh}{m_\varphi c^2} \, .
\end{equation}		
Fig.~\ref{fig:fDM} presents the shape of both velocity and frequency DM distribution. In particular, it is interesting that the shape of the frequency distribution is highly asymmetric because of the dispersion relation from Eq. (\ref{eq:ff}) and has a lower cut-off at the frequency $m_\varphi c^2/h$. This feature is particulary interesting to identify DM in the power spectrum of an experiment.

\subsection{Modeling of the scalar field}
In general, for a given scalar field mass $m_\varphi$, the scalar field can be written as \cite{Foster2018}
\begin{equation}\label{eq:varphi}
	\varphi(t) = \frac{\sqrt{\frac{4\pi G \rho_\mathrm{DM}}{c^2}}}{\frac{\hbar m_\varphi}{c^2}} \sum_{j=1}^{N_j}\alpha_j \sqrt{f_\mathrm{DM}(f_j)\Delta f}\cos\left[\omega_j t + \phi_j\right] \, ,
\end{equation}
where 
\begin{equation}
	\omega_j = 2\pi f_j = \frac{m_\varphi c^2}{\hbar}\left(1+\frac{v_j^2}{2 c^2}\right) \, .
\end{equation}
The amplitudes $\alpha_j$ follow a Rayleigh distribution \cite{Foster2018} while the phases $\phi_j$ are uniformly distributed, i.e.
\begin{subequations}\label{eq:priors}
	\begin{align}
		P[\phi_j] &= \frac{1}{2\pi} \hspace{5mm} \textrm{for}\hspace{5mm} 0\leq\phi_j\leq2\pi \, , \label{eq:priorphi}\\
		P[\alpha_j]&=\alpha_j e^{-\alpha_j^2/2}\, \label{eq:prioralpha}.
	\end{align}
\end{subequations}

The number of terms involved in the sum depends on the frequency resolution of the experiment $\Delta f=\frac{1}{T_\mathrm{exp}}$ and of the typical width of the frequency distribution. As can be noticed from Fig.~\ref{fig:fDM}, the full width half max (FWHM) of the frequency distribution, a good estimator of its width, is given by $\sim 10^{-6}m_\varphi c^2/h$. In practice, we use a sampling of the DM frequency distribution that covers $a$ FWHM starting at the cut-off frequency. In other words, the frequencies $f_j$ included in the sum from Eq. (\ref{eq:varphi}) are the Fourier frequencies (i.e. $f_j=j\Delta_f = j f_s/N$ with $f_s$ the sampling frequency and $N$ the number of measurements) contained in the range
\begin{equation}\label{eq:rangef}
	\left[\frac{m_\varphi c^2}{h},\frac{m_\varphi c^2}{h}\left(1+ a \times 10^{-6}\right)\right] \, ,
\end{equation}
where in practice we use $a=3$. The frequency region covered by this sampling is indicated by the green shaded area in Fig.~\ref{fig:fDM}.

The energy density for a scalar field is given by
\begin{equation}
	\rho_\varphi = \frac{c^2}{8\pi G}\left[\dot \varphi^2 + \frac{c^4m^2}{\hbar^2}\varphi^2\right]\, .
\end{equation}
For the scalar field from Eq.~(\ref{eq:varphi}), this quantity is a stochastic quantity. We can perform an ensemble average of the energy density for the scalar field using the distribution from Eqs.~(\ref{eq:priors}) to demonstrate that average energy density for the scalar field is the local DM energy density, i.e. $\left<\rho_\varphi\right>=\rho_\mathrm{DM}$.

\subsection{Modeling of the phase measurements}
Eq.~(4) from the main part of the paper gives the relationship between the phase measurement and the scalar field. If we take into account the fact that the scalar field has several frequencies (see Eq.~(\ref{eq:varphi})) and taking into account only the contribution from $d_e$ and $d_{m_e}$, the phase measurements are modeled as
\begin{align}
	\Delta\Phi(t)= \omega_0 T_0 &+\sum_j \alpha_j \left(d_e \tilde A_j+ d_{m_e}\bar A_j \right) \cos \left(\omega_j t +\phi_j+ \tilde \phi_j\right) \, ,
\end{align}
with
\begin{subequations}\label{eq:Aj}
	\begin{align}
	    \tilde A_j&=\frac{\sqrt{\frac{4\pi G \rho_\mathrm{DM}}{c^2}}}{\frac{\hbar m_\varphi}{c^2}}\sqrt{f_\mathrm{DM}(f_j)\Delta f} \sqrt{K_e^2(\omega_j)+L^2(\omega_j)} \, ,\\
		\bar   A_j&=\frac{\sqrt{\frac{4\pi G \rho_\mathrm{DM}}{c^2}}}{\frac{\hbar m_\varphi}{c^2}}\sqrt{f_\mathrm{DM}(f_j)\Delta f} \sqrt{K_{m_e}^2(\omega_j)+L^2(\omega_j)} \, ,
	\end{align}
\end{subequations}
where the functions $K(\omega)$ and $L(\omega)$ are obtained from equation (5) of the main part of the paper:
\begin{subequations}
	\begin{align}
		K_e(\omega)&=2\frac{\omega_0}{\omega}\sin\left(\omega\frac{T_0}{2}\right)\left[\alpha(\omega)-2\frac{\omega_0}{n_0}\frac{\partial n}{\partial \omega}\right]\, , \\
		K_{m_e}(\omega)&=2\frac{\omega_0}{\omega}\sin\left(\omega\frac{T_0}{2}\right)\left[\alpha(\omega)-\frac{3}{2}\frac{\omega_0}{n_0}\frac{\partial n}{\partial \omega}\right]\, , \\
		L(\omega)&=2\frac{\omega_0}{\omega}\sin\left(\omega\frac{T_0}{2}\right)\beta(\omega)\, ,
	\end{align}
\end{subequations}
and $\tilde\phi_j$ are constants, such that $\bar \phi_j= \phi_j+\tilde\phi_j$ are also uniformly distributed.

In this analysis, we use a ``maximum reach approach'' which means that we are considering $d_e$ and $d_{m_e}$ independently in two independent analysis where we fix one of these parameters to 0. We can then write the signal that is used in our data analysis as
\begin{equation}\label{eq:signal}
	s\left(t,\gamma,\left\{\alpha_j\right\},\left\{\varphi_j\right\}\right) = \omega_0 T_0 + \gamma \sum_{j=1}^{N_j} \alpha_j A_j \cos\left[\omega_j t +\bar \phi_j\right] \, ,
\end{equation}
where we use $\gamma=d_e$ and $A_j=\tilde A_j$ when we consider the coupling parameter to electromagnetism and $\gamma=d_{m_e}$ with $A_j=\bar A_j$ when we consider the coupling parameter to the electron mass.

\subsection{Fourier transform}
In order to infer the value of $\gamma$, the linear combination of DM coupling constants, we choose to analyse the data using a Bayesian inference method on a discrete Fourier transform (DFT) of our full dataset. In this section, we briefly remind our notation for DFT and useful relations. We follow closely the Appendix of \cite{Derevianko2016}.

We have one set of $N$ measurements sampled at the frequency $f_s=\frac{1}{\Delta t}$ over a period $T_\mathrm{exp}=N/f_s$ with colored noise characterized by its PSD from Eq.~(\ref{eq:PSD_appendix}), i.e. the dataset can be described as a set of $\left\{(t_i,d_i)\right\}$ measurements caracterized by a covariance matrix $C_{ij}$.

For any time depend function $x(t)$, we will write the value of $x(t)$ at the $l^\mathrm{th}$ sampling time by $x_l=x(l\Delta t)$ where $l\in \left[0, 1, \dots N-1\right]$. A tilde quantity will denote the discrete Fourrier Transform (DFT) of a quantity 
	\begin{equation}
		\tilde x_k = \sum_{l=0}^{N-1} e^{-2\pi i\frac{kl}{N} }x_l \,  , 
	\end{equation} 
	where  $k\in \left[0,1,\dots,N-1\right]$. Note that $\tilde x_k$ corresponds to the frequency $f_k=k f_s/N$, it is periodic $\tilde x_{k+N}=\tilde x_k$ and if $x$ is a real signal, it is symmetric $\tilde x_{N-k}=\bar{\tilde x}_k$ where a bar denotes the complex conjugate.  
	In vectorial notation, on can write the last equation as
	\begin{equation}\label{eq:DFT}
		\bm{\tilde x} = \sqrt{N} \bm U \bm x\, ,
	\end{equation}
	where we introduced the rotation matrix $\bm U$ whose components are $U_{kl}=\mathrm{exp}(-2\pi i \frac{kl}{N})/\sqrt{N}$. This matrix $\bm U$ is symmetric and unitary ($\bm U \cdot\bm U^\dagger=\bm U^\dagger\cdot \bm U=\bm \delta$ with $\bm \delta$ the identity).

Let us introduce the noise time series $\bm n$ which has a vanishing expectation $E\left[n_i\right]=0$. The noise covariance matrix then is given by $C_{ij}=E\left[n_i \bar n_j\right]$. A simple calculation shows that the covariance matrix of the DFT $\tilde {\bm n}$ is given by $\tilde{\bm C}= N \bm U\cdot\bm C\cdot \bm U^\dagger $ such that
\begin{equation}\label{eq:C-1}
	\bm C^{-1} = N \bm U^\dagger \cdot\tilde{\bm C}^{-1} \cdot \bm U \, .
\end{equation}
$\tilde {\bm C}$ is known as the \textbf{two-sided} PSD matrix, which for a stationnary process is diagonal \cite{Derevianko2016}. We can introduce the two-sided PSD by $\tilde C_{ij}=Nf_s\delta_{ij}S_j$ or in other words
\begin{equation}
	S_j=S(f_j)=\frac{E[\left|\tilde n_j\right|^2]}{N f_s}\, .
\end{equation}

\subsection{Bayesian inference}
Working in the context of Bayesian inference, we will use a Gaussian likelihood (i.e. the probability to get the dataset given the model and some model parameters) with a colored noise, which writes
	 \begin{equation}
	 	 \mathcal L = P\left[\bm d| \gamma, \bm \alpha , \bar{\bm \phi} \right] = \frac{1}{\sqrt{\mathrm{det}(2\pi\bm C)}}\mathrm{exp}\left(-\frac{1}{2} (\bm d-\bm s)^\dagger \cdot \bm C^{-1} \cdot (\bm d-\bm s)\right)\, ,
	 \end{equation}
	 where $\bm d = (d_0, d_1, \dots, d_{N-1})$ is the vector of data and $\bm s(\bm p) = (s(t_0, \bm p), s(t_1, \bm p), \dots, s(t_{N-1}, \bm p))$ is the model given by Eq.~(\ref{eq:signal}) while $\bm C$ is the noise covariance matrix. Note that the model depends on one coupling parameter $\gamma$ and on a set of amplitudes $\alpha_j$ and of phases $\bar \phi_j$.
	
Using the DFT of the data and of the signal such as introduced in Eq.~(\ref{eq:DFT}) as well as Eq.~(\ref{eq:C-1}), the likelihood becomes
\begin{subequations}\label{eq:like_transf}
	\begin{align}
		\mathcal L  & = \frac{1}{\prod_j (2\pi S_j f_s)^{1/2}} \mathrm{exp}\left(-\frac{1}{2} (\bm {\tilde d}-\bm {\tilde s})^T \cdot \bm C^{-1} \cdot (\bm {\tilde d}-\bm {\tilde s})\right)= \frac{1}{\prod_j (2\pi S_j f_s)^{1/2}}  \mathrm{exp}\left(-\frac{1}{2}\sum_{k=0}^{N-1} \frac{\left|\tilde d_k-\tilde s_k\right|^2}{\tilde C_{kk}}\right)\, \\
		&= \frac{1}{\prod_j (2\pi S_j f_s)^{1/2}}  \mathrm{exp}\left(-\sum_{k=0}^{\lf N/2\rf }\beta_k \frac{\left|\tilde d_k-\tilde s_k\right|^2}{\tilde C_{kk}}\right)\, ,
	\end{align}
	\label{eq:Likelihood}
\end{subequations}	
where $\beta_k=1/2$ for $k=0$, $\beta_k=1/2$ for $k=N/2$ when $N$ is even and $\beta_k=1$ otherwise. In our case, we will not consider the 0 and higher frequency in our analysis so $\beta_k=1$ (but see the Appendix of \cite{Derevianko2016} for a general case). In the end, the log-likelihood writes
\begin{equation}\label{eq:loglike}
	- \ln \mathcal L(\bm d| \gamma,\bm \alpha,\bm \phi) = \sum_{k=0}^{\lf N/2 \rf} -\ln \mathcal L_k =  \sum_{k=0}^{\lf N/2 \rf} \tilde \chi_k^2 + \ln \left(S_k f_s\right) \, ,
\end{equation}
with
\begin{equation}\label{eq:tildechi2}
	\tilde \chi_k^2 = \frac{\left|\tilde d_k-\tilde s_k\right|^2}{\tilde C_{kk}} = \frac{\left|\tilde d_k-\tilde s_k\right|^2}{N f_s S_k} \, .
\end{equation}

Since the signal is modeled as a sum of harmonic components at Fourrier frequencies (see Eq.~(\ref{eq:signal}) where $f_j$ have been choosen as $f_j=\frac{j}{N} f_s$), the DFT of the signal can easily be computed
\begin{equation}
	\tilde s_k = \frac{\gamma N A_k \alpha_k}{2}e^{i\bar \phi_k} \quad \textrm{for}\quad  k > 0 \, .
\end{equation}
	
The $\tilde\chi_k^2$ that appears in the expression of the likelihood from Eq.~(\ref{eq:tildechi2}) becomes
	\begin{equation}
		\tilde \chi^2_k=\frac{\left|\tilde d_k-\tilde s_k\right|^2}{N f_s S_k} = \frac{\left|\tilde d_k\right|^2}{N f_s S_k}+\frac{\left|\tilde s_k\right|^2}{N f_s S_k}-\frac{2}{N f_s S_k}\mathcal Re\left[\overline {\tilde d_k} \tilde s_k\right]\, ,
	\end{equation}
	where $\mathcal Re[x]$ denotes the real part of $x$. If we introduce $\theta_k$ such that $\tilde d_k=|\tilde d_k|e^{i\theta_k}$ then
	\begin{equation}
		\tilde \chi^2_k=\frac{\left|\tilde d_k\right|^2}{N f_s S_k}+\frac{\gamma^2 N A_k^2\alpha_k^2}{4 f_s S_k}-\frac{\gamma  A_k\alpha_k|\tilde d_k|}{f_s S_k}\cos(\bar\phi_k+\theta_k) \, .
		\label{eq:chi2}
	\end{equation}

The likelihood from Eq.~(\ref{eq:loglike}) depends on a large number of parameters ($\gamma$, $\bar \phi_j$ and $\alpha_j$) making it hard to sample efficiently. Since we are not interested in the estimates of $\alpha_j$ and $\bar\phi_j$, we can marginalize the likelihood over these parameters. The first step consists in integrating on the random phases $\bar\phi_j$ such that the likelihood marginalized over the phases $\mathcal L (\bm d| \gamma,\bm \alpha)=\int d\bar {\bm \phi} \mathcal L (\bm d| \gamma,\bm \alpha,\bar {\bm \phi}) P\left[\bar {\bm \phi}\right]$, where the last term is the prior from Eq.~(\ref{eq:priorphi}). One can treat the frequencies independently and 
	\begin{align}
		\mathcal L_k (\bm d| \gamma, \alpha_k) &= \frac{1}{2\pi}\int_0^{2\pi} d\bar \phi_k \frac{1}{2\pi f_s S_k}e^{-\tilde\chi_k^2}\nonumber \\ 
	  &=\frac{1}{2\pi f_s S_k} e^{-\frac{\left|\tilde d_k\right|^2}{N f_s S_k}} e^{-\frac{N \gamma^2A_k^2\alpha_k^2}{4 f_s S_k}} I_0\left(\frac{\gamma  A_k\alpha_k\left|\tilde d_k\right|}{f_s S_k}\right)\, ,
	\end{align}
	with $I_0(x)$ the Bessel function.

	The second step is a marginalization over $\alpha_j$. Once again, we can treat each frequency independantly such that $\mathcal L_k (\bm d| \gamma)=\int_0^\infty d\alpha_k \mathcal L_k (\bm d| \gamma, \alpha_k) P\left[\alpha_k\right]$, where the last term is the Rayleigh prior from Eq.~(\ref{eq:prioralpha}), i.e.
	\begin{equation}
		 \mathcal L_k (\bm d| \gamma)=  \frac{1}{2\pi f_s S_k} e^{-\frac{\left|\tilde d_k\right|^2}{Nf_s S_k}} \int_0^\infty d{\alpha_k} \alpha_k  e^{-\alpha_k^2 \left(\frac{N \gamma^2A_k^2}{4 f_s S_k}+\frac{1}{2}\right)} I_0\left(\frac{\gamma  A_k\alpha_k\left|\tilde d_k\right|}{f_s S_k}\right)\, .	
	 \end{equation}
Fortunately, this expression is analytical since
$$
	\int_0^\infty dx \, x \, e^{-ax^2}I_0(b x)=\frac{e^{b^2/4a}}{2a} \, ,
$$
and leads to an expression of the marginalized likelihood given by
\begin{equation}\label{eq:posterior_notwhite}
		 \mathcal L_k \left(\bm d | \gamma \right)=   \frac{1}{2\pi f_s S_k}  \frac{1}{1+\frac{N \gamma^2A_k^2}{2 f_s S_k}} \exp \left( \frac{-\frac{\left|\tilde d_k\right|^2}{Nf_sS_k}}{1+\frac{N \gamma^2A_k^2}{2 f_s S_k}} \right)\, .
\end{equation} 

Finally the posterior distribution on $\gamma$ $\mathcal P(\gamma | \bm d)$ marginalized over all other parameters is given by the Bayes theorem and is $\mathcal P(\gamma | \bm d)\propto \mathcal L(\bm d|\gamma ) P\left[\gamma\right]$ where the last term is the prior on $\gamma$ that is choosen as flat. In the end, the posterior is given by \cite{Derevianko2016,Foster2018,Centers2019}
	\begin{equation}\label{eq:posterior_final}
			-\ln  \mathcal P(\gamma | \bm d) = \mathrm{cst} + \sum_{k=1}^{\lf N/2\rf} \frac{\frac{\left|\tilde d_k\right|^2}{N f_s S_k}}{1+\frac{N \gamma^2A_k^2}{2 f_s S_k}} +\ln \left(1+\frac{N \gamma^2A_k^2}{2 f_s S_k}\right)\, .
		\end{equation}

The 95\% upper value for $\gamma$ is determined from this posterior distribution by solving for
\begin{equation}\label{eq:gamm95}
	\int_{-\gamma_{95\%}}^{\gamma_{95\%}}\mathcal P(\gamma | \bm d)d\gamma =2 \int_{0}^{\gamma_{95\%}}\mathcal P(\gamma | \bm d)d\gamma=0.95\, .
\end{equation}

\subsection{Summary of the data analysis in practice}
From the raw measurements $\bm d$, we compute the DFT and compute the $\left|\tilde d_k\right|^2$ values. Then, for a given mass of the scalar field $m_\varphi$, we compute the range over which the DM frequency distribution is non-negligible, i.e. we use Eq.~(\ref{eq:rangef}) with $a=3$. We determine the Fourier frequencies $f_j= \frac{j}{N}f_s$ that are contained in this frequency range. For all these frequencies, we evaluate the values of $A_j$ that are given by Eq.~(\ref{eq:Aj}) and the two-sided PSD $S_k$ is provided by Eq.~(\ref{eq:PSD_appendix}). We can then evaluate the posterior $\mathcal P(\gamma | \bm d)$ using Eq.~(\ref{eq:posterior_final}) and compute the 95\% upper limit by solving numerically for Eq.~(\ref{eq:gamm95}). We iterate this procedure for all masses corresponding to the $[10,200]$~kHz frequency range.

\subsection{Data analysis in the Earth relaxion halo model}
In this model the density of DM on the Earth's surface is much larger than in the galactic DM model (see supplementary figure 2 of \cite{Banerjee2020}) and we simply include this as an additional frequency dependent pre-factor in eqs. (\ref{eq:Aj}). The velocity distribution of DM is also different but the corresponding coherence times are much larger than for the galactic DM distribution. As a consequence the width of the corresponding frequency distribution is smaller than the RAM limited frequency resolution of our DFT ($\sim 3$~mHz) and we use a single frequency $\omega_j$ in the sum of (\ref{eq:varphi}). The rest of the procedure is identical to the galactic DM case described above. Note that we still take the probability distribution of $\alpha_j$ and $\phi_j$ into account and marginalise over them. As a consequence the factor $\sim$10 sensitivity loss pointed out in \cite{Centers2019} is accounted for, contrary to e.g. \cite{Aharony2019,Antypas2019}. Note that one could assume that the relaxation halo is condensed in the ground state (see e.g. the short discussion around eqs. (4) and (5) of \cite{Banerjee2019}) in which case there is no stochastic energy distribution and the analysis is significantly simplified.

\end{document}